# New Phase Space Formulations and Quantum Dynamics Approaches

Xin He[1, †], Baihua Wu[1, †], Youhao Shang[1], Bingqi Li[1], Xiangsong Cheng[1], and Jian Liu[1, *]

[1]Beijing National Laboratory for Molecular Sciences, Institute of Theoretical and Computational Chemistry, College of Chemistry and Molecular Engineering, Peking University, Beijing 100871, China.

[†] X. H. and B. W. contributed equally.

[*] **Corresponding Author:** Email: jianliupku@pku.edu.cn

**Abstract**

We report recent progress on the phase space formulation of quantum mechanics with coordinate-momentum variables, focusing more on new theory of (weighted) constraint coordinate-momentum phase space for discrete-variable quantum systems. This leads to a general coordinate-momentum phase space formulation of composite quantum systems, where conventional representations on infinite phase space are employed for continuous variables. It is convenient to utilize (weighted) constraint coordinate-momentum phase space for representing the quantum state and describing nonclassical features. Various numerical tests demonstrate that new trajectory-based quantum dynamics approaches derived from the (weighted) constraint phase space representation are useful and practical for describing dynamical processes of composite quantum systems in gas phase as well as in condensed phase.

**Graphical Abstract**

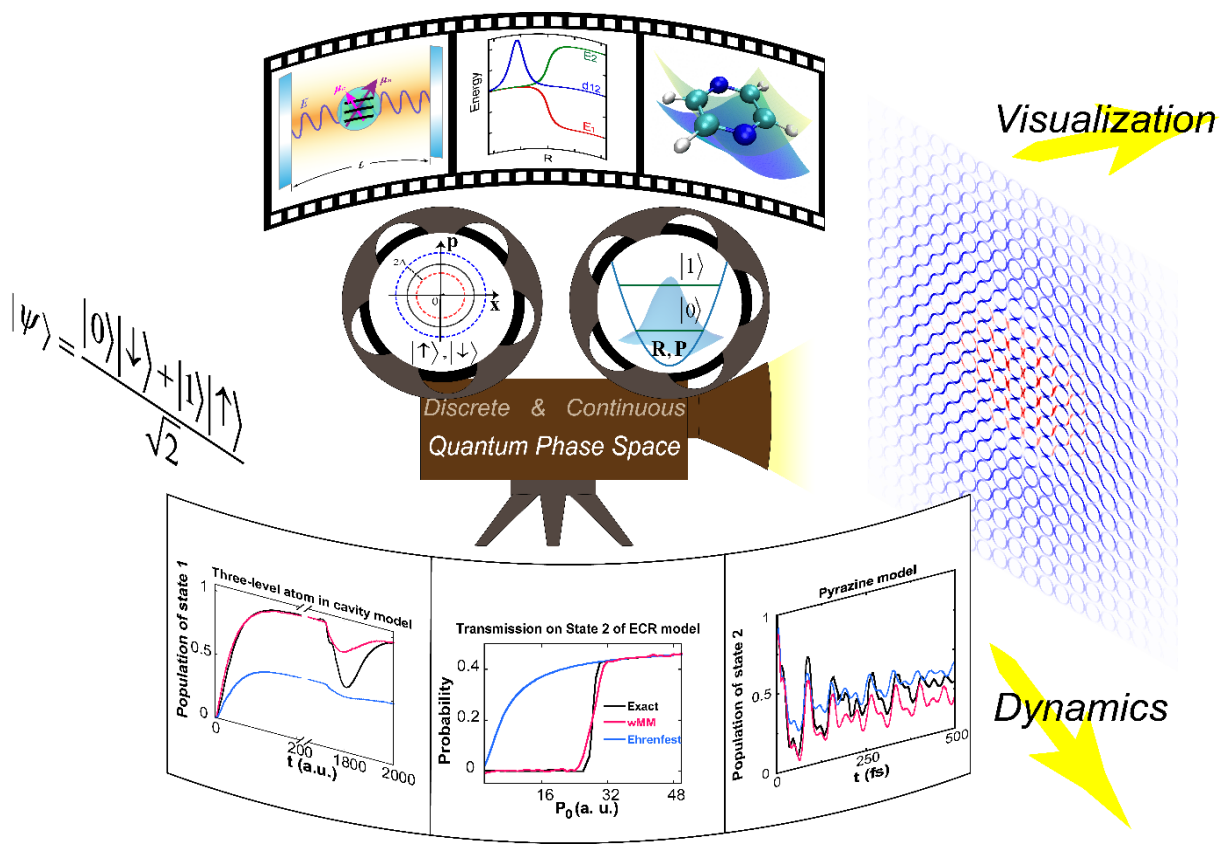

**Schematic representation of dynamics of composite quantum systems *via* phase space formulations.**

# 1. INTRODUCTION

Phase space with coordinate-momentum variables is a fundamental concept and offers a convenient tool to describe statistics as well as dynamics in classical mechanics. In comparison to other equivalent interpretations of quantum mechanics, phase space formulations offer more insight and understanding between quantum and classical counterpart concepts, which are widely used in chemical and biological dynamics and spectroscopy[1-60], quantum optics[51, 61-70], cryogenic physics/chemistry[71-75], quantum information and computation[76-87], and so forth.

Phase space formulations of quantum mechanics have been developed since two important pioneering works, the Weyl transform in 1927, of which the original formulation converted a Hamiltonian on classical phase space into a quantum mechanical operator[88], and the Wigner function in 1932 that in principle depicts the inverse transform although a pure state was used for demonstration[89].  The most essential element is the one-to-one correspondence mapping between quantum operators and classical functions often defined on a smooth manifold, namely, phase space.   Because of the commutation relation of conjugate operators, the mapping is not unique in quantum mechanics[90, 91].

When infinite phase space is employed for a continuous-variable quantum system, most phase space formulations can be described by Cohen's generalized form[92] in 1966.  Quantum dynamics with phase space variables is expressed by the Moyal or Moyal-like bracket as first proposed by Groenewold[93] in 1946 and Moyal[94] in 1949.  The Wigner and Husimi representations are most often used for the continuous-variable system. When the Moyal bracket is approximated by the Poisson bracket in the Wigner phase space expression of the quantum Liouville theorem, which was also derived as the linearized semiclassical initial value representation (LSC-IVR) or classical Wigner model[4-8, 11, 20, 23] for the quantum correlation function, it reproduces exact quantum correlation functions even of nonlinear operators (i.e., nonlinear functions of the coordinate or momentum operator) in the harmonic or classical limit. The truncated Wigner approximation[74] with the time-dependent generalization of the Bopp representation[90, 95] is similar to the LSC-IVR, but the former requests more demanding evaluation of the stability matrix elements along the trajectory when nonlinear operators are involved in the correlation function.  Liu and Miller [15] suggest a practical way to implement the imaginary time path integral treatment of the Boltzmann density operator in the LSC-IVR for general molecular systems that often contain imaginary frequencies.  Its recent application illustrates that quantum dynamical effects play a critical role in reproducing the peaks in the intermediate region between the librational and bending bands, those between the bending and stretching bands, and the double-peak in the stretching band in the experimental isotropic Raman spectrum of liquid water[19] (as shown in Figure 1).  In addition that more advanced versions of SC-IVR[96-99] are capable of improving over the LSC-IVR, in Reference [11] we first employed the quantum Liouville theorem in the phase space formulation to develop trajectory-based approaches to satisfy the two fundamental criteria: conservation of the quantum Boltzmann distribution for the thermal equilibrium system and being exact for any quantum thermal correlation functions in the classical and harmonic limits. Such trajectory-based approaches can in principle be further improved by higher order corrections of the exact series expansion of the phase space propagator as demonstrated in Reference [44]. More progress along this line can be found in References [37-47]. (Figure 2 shows molecular vibrational spectra produced by the new phase space quantum dynamics methods.)

Phase space representations of a finite discrete $F$-state quantum system were first independently described by Stratonovich[100] in 1956, Feynman[101] in 1987, and Wootters[102] in 1987. Further developments of Stratonovich's formulation have focused on an SU(2) or SU($F$) structure of phase space[103-117], while those on the construction of a discrete phase space are described in References [78, 118-126]. Other than the 2-state (or spin 1/2) system, the exact equations of motion (EOMs) of phase variables (expressed by the Moyal-like bracket) involved in these approaches for the finite discrete multi-state system are often tedious and numerically unfavourable[109, 127-132]. (See Appendix 3 of the Supporting Information for more discussion.) Recent theoretical progress on exactly mapping the finite discrete $F$-state quantum system onto *constraint* coordinate-momentum phase space suggests that there exists a novel unified framework to derive comprehensive exact mapping Hamiltonians[44, 57, 133, 134], of which the quantum EOMs of mapping coordinate-momentum variables are simply linear[44, 57, 133-136].

The unified mapping formulation on coordinate-momentum phase space[44, 57, 133-136] then offers a useful tool to treat dynamics of a composite quantum system, in which both continuous and finite discrete degrees of freedom (DOFs) are involved and coupled with one another. Because a typical molecular system has vibrational, rotational, and translational motion, it is often much more convenient to employ continuous coordinate space rather than Hilbert space with dense states to describe the nuclear DOFs. On the other hand, the energy gap between different electronic states of interest is often significantly larger such that the (adiabatic or diabatic) state representation is more useful to depict the electronic DOFs. It is evident that a general description of the molecular system leads to a composite quantum system, especially in the nonadiabatic region[137-154]. A comprehensive version of the Meyer-Miller mapping Hamiltonian model[155, 156] can rigorously be formulated in the general coordinate-momentum phase space formulation[44, 57, 133-136].

In the Focus Article we focus on novel developments on the phase space formulation of quantum mechanics with coordinate-momentum variables for discrete-variable systems as well as for composite systems[44, 57, 133-136]. In Section 2 we first review the general coordinate-momentum phase space formulation, where infinite space is used for describing continuous variables and constraint space is employed for mapping discrete variables. We then propose a weighted constraint phase space representation that is also an exact formulation for mapping discrete-variable quantum systems. Section 3 demonstrates several examples and discusses implications of the (weighted) constraint coordinate-momentum phase space for studying and illustrating discrete-variable or composite quantum systems. When we use the weighted constraint phase space representation for mapping composite quantum systems, the mapping Hamiltonian (we use the Meyer-Miller mapping Hamiltonian for demonstration throughout the article, albeit that other mapping Hamiltonians are also available[57, 58, 133, 134]) yields a novel trajectory-based approximate approach for composite systems. Such a new method satisfies the frozen nuclei limit [i.e., the dynamics reproduces the exact evolution when only finite discrete (electronic) DOFs are involved]. In Section 4 the performance of new trajectory-based quantum dynamics approaches on (weighted) constraint phase space is extensively tested for a few typical benchmark composite systems in gas phase as well as in condensed phase. Finally, conclusion remarks are presented in Section 5.

## 2. GENERAL COORDINATE-MOMENTUM PHASE SPACE FORMULATION OF QUANTUM MECHANICS

Consider a (molecular) system with $N$ continuous (nuclear) DOFs and $F$ discrete (electronic) states, of which the Hamiltonian reads

$$\hat{H} = \sum_{n,m=1}^{F} H_{nm}\left(\hat{\mathbf{R}}, \hat{\mathbf{P}}\right) |n\rangle\langle m| = \sum_{n,m=1}^{F} \left[ \frac{1}{2} \hat{\mathbf{P}}^{\mathrm{T}} \mathbf{M}^{-1} \hat{\mathbf{P}} \delta_{nm} + V_{nm}\left(\hat{\mathbf{R}}\right) \right] |n\rangle\langle m| \quad , \tag{1}$$

where $\mathbf{R}$ and $\mathbf{P}$ are the nuclear coordinate and momentum variables, respectively, $\mathbf{M}$ is the diagonal mass matrix, and the $F$ states form an orthonormal complete basis set, that is,

$$\langle m|n\rangle = \delta_{mn}, \quad \hat{I}_{\mathrm{ele}} = \sum_{n=1}^{F} |n\rangle\langle n| \quad . \tag{2}$$

$\hat{I}_{\mathrm{ele}}$ and $\hat{I}_{\mathrm{nuc}}$ stand for the identity operator of the discrete (electronic) DOFs and that of the continuous (nuclear) DOFs. For simplicity, eq (1) employs the (electronically) diabatic representation, where the Hermitian potential matrix $\mathbf{V}(\mathbf{R})$ is a function of only the coordinate vector. (In applications $\mathbf{V}(\mathbf{R})$ is often a real symmetric matrix.) More discussion on the adiabatic representation of discrete (electronic) DOFs is available in Section-4.1.

The unified formulation of mapping phase space with coordinate-momentum variables offers a useful exact approach to describe the composite system. The trace of a product of two quantum operators is expressed as an integral of two functions on mapping phase space, that is,

$$\mathrm{Tr}_{n,e}\left[\hat{A}\hat{B}\right] = \int d\boldsymbol{\mu}_{\mathrm{nuc}}(\mathbf{R},\mathbf{P}) \int_{\mathcal{S}(\mathbf{x},\mathbf{p})} d\boldsymbol{\mu}_{\mathrm{ele}}(\mathbf{x},\mathbf{p}) A_C(\mathbf{R},\mathbf{P};\mathbf{x},\mathbf{p})\tilde{B}_C(\mathbf{R},\mathbf{P};\mathbf{x},\mathbf{p}) \quad (3)$$

with

$$A_C(\mathbf{R},\mathbf{P};\mathbf{x},\mathbf{p}) = \mathrm{Tr}_{n,e}\left[\hat{A}\hat{K}_{\mathrm{nuc}}(\mathbf{R},\mathbf{P})\otimes\hat{K}_{\mathrm{ele}}(\mathbf{x},\mathbf{p})\right] \quad , \quad (4)$$

$$\tilde{B}_C(\mathbf{R},\mathbf{P};\mathbf{x},\mathbf{p}) = \mathrm{Tr}_{n,e}\left[\hat{K}_{\mathrm{nuc}}^{-1}(\mathbf{R},\mathbf{P})\otimes\hat{K}_{\mathrm{ele}}^{-1}(\mathbf{x},\mathbf{p})\hat{B}\right] \quad , \quad (5)$$

$d\boldsymbol{\mu}_{\mathrm{nuc}}(\mathbf{R},\mathbf{P}) = (2\pi\hbar)^{-N} d\mathbf{R}d\mathbf{P}$ and $d\boldsymbol{\mu}_{\mathrm{ele}}(\mathbf{x},\mathbf{p}) = F d\mathbf{x}d\mathbf{p}$ as the integration measure on nuclear phase space and that on electronic phase space, respectively, and $\mathrm{Tr}_{n,e}$ represents the trace over the corresponding nuclear and electronic Hilbert space. The integral over the mapping phase space variables for the finite discrete (electronic) DOFs in eq (3) is performed as

$$\int_{\mathcal{S}(\mathbf{x},\mathbf{p})} F d\mathbf{x}d\mathbf{p}\, g(\mathbf{x},\mathbf{p}) = \int F d\mathbf{x}d\mathbf{p}\frac{1}{\Omega}\mathcal{S}(\mathbf{x},\mathbf{p})g(\mathbf{x},\mathbf{p}) = \int F d\mathbf{x}d\mathbf{p}\,\overline{\mathcal{S}}(\mathbf{x},\mathbf{p})g(\mathbf{x},\mathbf{p}) \quad , \quad (6)$$

where the area of constraint space $\mathcal{S}(\mathbf{x},\mathbf{p})$

$$\Omega = \int d\mathbf{x}d\mathbf{p}\,\mathcal{S}(\mathbf{x},\mathbf{p}) \quad (7)$$

is the normalization constant, and $\overline{\mathcal{S}}(\mathbf{x},\mathbf{p})$ is the normalized constraint space.

The normalization of the (inverse) mapping kernel reads

$$\mathrm{Tr}_n\left[\hat{K}_{\mathrm{nuc}}(\mathbf{R},\mathbf{P})\right] = \mathrm{Tr}_n\left[\hat{K}_{\mathrm{nuc}}^{-1}(\mathbf{R},\mathbf{P})\right] = 1 \quad (8)$$

$$\mathrm{Tr}_e\left[\hat{K}_{\mathrm{ele}}(\mathbf{x},\mathbf{p})\right] = \mathrm{Tr}_e\left[\hat{K}_{\mathrm{ele}}^{-1}(\mathbf{x},\mathbf{p})\right] = 1 \quad (9)$$

and

$$\int d\boldsymbol{\mu}_{\mathrm{nuc}}(\mathbf{R},\mathbf{P})\,\hat{K}_{\mathrm{nuc}}(\mathbf{R},\mathbf{P}) = \int d\boldsymbol{\mu}_{\mathrm{nuc}}(\mathbf{R},\mathbf{P})\,\hat{K}_{\mathrm{nuc}}^{-1}(\mathbf{R},\mathbf{P}) = \hat{I}_{\mathrm{nuc}} \quad (10)$$

$$\int_{\mathcal{S}(\mathbf{x},\mathbf{p})} d\boldsymbol{\mu}_{\mathrm{ele}}(\mathbf{x},\mathbf{p})\hat{K}_{\mathrm{ele}}(\mathbf{x},\mathbf{p}) = \int_{\mathcal{S}(\mathbf{x},\mathbf{p})} d\boldsymbol{\mu}_{\mathrm{ele}}(\mathbf{x},\mathbf{p})\,\hat{K}_{\mathrm{ele}}^{-1}(\mathbf{x},\mathbf{p}) = \hat{I}_{\mathrm{ele}} \quad . \quad (11)$$

The one-to-one correspondence mapping from phase space function $A_C(\mathbf{R},\mathbf{P};\mathbf{x},\mathbf{p})$ or $\tilde{B}_C(\mathbf{R},\mathbf{P};\mathbf{x},\mathbf{p})$ of Eq (4) back to operator $\hat{A}$ or $\hat{B}$ is

$$\begin{aligned} \hat{A} &= \int d\boldsymbol{\mu}_{\text{nuc}}(\mathbf{R},\mathbf{P}) \int_{\mathcal{S}(\mathbf{x},\mathbf{p})} d\boldsymbol{\mu}_{\text{ele}}(\mathbf{x},\mathbf{p}) A_C(\mathbf{R},\mathbf{P};\mathbf{x},\mathbf{p}) \hat{K}_{\text{nuc}}^{-1}(\mathbf{R},\mathbf{P}) \otimes \hat{K}_{\text{ele}}^{-1}(\mathbf{x},\mathbf{p}) \\ \hat{B} &= \int d\boldsymbol{\mu}_{\text{nuc}}(\mathbf{R},\mathbf{P}) \int_{\mathcal{S}(\mathbf{x},\mathbf{p})} d\boldsymbol{\mu}_{\text{ele}}(\mathbf{x},\mathbf{p}) \tilde{B}_C(\mathbf{R},\mathbf{P};\mathbf{x},\mathbf{p}) \hat{K}_{\text{nuc}}(\mathbf{R},\mathbf{P}) \otimes \hat{K}_{\text{ele}}(\mathbf{x},\mathbf{p}) \end{aligned} \quad (12)$$

The nuclear or electronic kernel should satisfy five criteria, namely, linearity, reality, standardization (normalization), traciality, and covariance[93, 94, 100, 115].

### 2.1 Mapping kernel for continuous (nuclear) degrees of freedom

The integrals for $(\mathbf{R},\mathbf{P})$ in eqs (3), (10), and (12) are over infinite (nuclear) phase space. The mapping kernel and its inverse for the nuclear DOFs are

$$\begin{aligned} \hat{K}_{\text{nuc}}(\mathbf{R},\mathbf{P}) &= \left(\frac{\hbar}{2\pi}\right)^N \int d\boldsymbol{\zeta} \int d\boldsymbol{\eta}\, e^{i\boldsymbol{\zeta}\cdot(\hat{\mathbf{R}}-\mathbf{R})+i\boldsymbol{\eta}\cdot(\hat{\mathbf{P}}-\mathbf{P})} f(\boldsymbol{\zeta},\boldsymbol{\eta}) \\ \hat{K}_{\text{nuc}}^{-1}(\mathbf{R},\mathbf{P}) &= \left(\frac{\hbar}{2\pi}\right)^N \int d\boldsymbol{\zeta} \int d\boldsymbol{\eta}\, e^{i\boldsymbol{\zeta}\cdot(\hat{\mathbf{R}}-\mathbf{R})+i\boldsymbol{\eta}\cdot(\hat{\mathbf{P}}-\mathbf{P})} \left[f(-\boldsymbol{\zeta},-\boldsymbol{\eta})\right]^{-1} \end{aligned} \quad , \quad (13)$$

where $f(\boldsymbol{\zeta},\boldsymbol{\eta})$ is a scalar function. For example, we have the Wigner function[89, 157]

$$f(\boldsymbol{\zeta},\boldsymbol{\eta}) = 1 \quad , \quad (14)$$

the Husimi function[158]

$$f(\boldsymbol{\zeta},\boldsymbol{\eta}) = \exp\left(-\frac{\boldsymbol{\zeta}^T \boldsymbol{\Gamma}^{-1} \boldsymbol{\zeta}}{4} - \frac{\hbar^2}{4}\boldsymbol{\eta}^T \boldsymbol{\Gamma} \boldsymbol{\eta}\right) \quad , \quad (15)$$

the anti-Husimi function

$$f(\boldsymbol{\zeta},\boldsymbol{\eta}) = \exp\left(\frac{\boldsymbol{\zeta}^T \boldsymbol{\Gamma}^{-1} \boldsymbol{\zeta}}{4} + \frac{\hbar^2}{4}\boldsymbol{\eta}^T \boldsymbol{\Gamma} \boldsymbol{\eta}\right) \quad , \quad (16)$$

the Glauber-Sudarshan P function[61, 62, 66] (with the characteristic frequency matrix $\boldsymbol{\omega}$ of the system)

$$f(\boldsymbol{\zeta},\boldsymbol{\eta}) = \exp\left[\frac{\hbar}{4}\boldsymbol{\zeta}^T \mathbf{M}^{-1/2}\boldsymbol{\omega}^{-1}\mathbf{M}^{-1/2}\boldsymbol{\zeta} + \frac{\hbar}{4}\boldsymbol{\eta}^T \mathbf{M}^{1/2}\boldsymbol{\omega}\mathbf{M}^{1/2}\boldsymbol{\eta}\right] \quad (17)$$

and its generalized versions[66], the Glauber Q function[159]

$$f(\boldsymbol{\zeta}, \boldsymbol{\eta}) = \exp\left[-\frac{\hbar}{4}\boldsymbol{\zeta}^T \mathbf{M}^{-1/2}\boldsymbol{\omega}^{-1}\mathbf{M}^{-1/2}\boldsymbol{\zeta} - \frac{\hbar}{4}\boldsymbol{\eta}^T \mathbf{M}^{1/2}\boldsymbol{\omega}\mathbf{M}^{1/2}\boldsymbol{\eta}\right] \quad , \tag{18}$$

the normal-antinormal ordered function[91]

$$f(\boldsymbol{\zeta}, \boldsymbol{\eta}) = \cosh\left[\frac{\hbar}{4}\boldsymbol{\zeta}^T \mathbf{M}^{-1/2}\boldsymbol{\omega}^{-1}\mathbf{M}^{-1/2}\boldsymbol{\zeta} + \frac{\hbar}{4}\boldsymbol{\eta}^T \mathbf{M}^{1/2}\boldsymbol{\omega}\mathbf{M}^{1/2}\boldsymbol{\eta}\right] \quad , \tag{19}$$

the Kirkwood antistandard-ordered function[160, 161]

$$f(\boldsymbol{\zeta}, \boldsymbol{\eta}) = e^{i\hbar \boldsymbol{\zeta}^T \boldsymbol{\eta}/2} \quad , \tag{20}$$

the Mehta standard-ordered function[162]

$$f(\boldsymbol{\zeta}, \boldsymbol{\eta}) = e^{-i\hbar \boldsymbol{\zeta}^T \boldsymbol{\eta}/2} \quad , \tag{21}$$

the Rivier function[163, 164]

$$f(\boldsymbol{\zeta}, \boldsymbol{\eta}) = \cos\left[\tfrac{1}{2}\hbar \boldsymbol{\zeta}^T \boldsymbol{\eta}\right] \quad , \tag{22}$$

and the distribution function of Born and Jordan[165]

$$f(\boldsymbol{\zeta}, \boldsymbol{\eta}) = \frac{\sin\left[\tfrac{1}{2}\hbar \boldsymbol{\zeta}^T \boldsymbol{\eta}\right]}{\tfrac{1}{2}\hbar \boldsymbol{\zeta}^T \boldsymbol{\eta}} \quad , \tag{23}$$

and so forth.

When operator $\hat{A}$ is a function of only the nuclear DOFs, its phase space function from Eq (4) and the dual function from eq (5) become

$$A_{\text{nuc}}(\mathbf{R}, \mathbf{P}) = \text{Tr}_n\left[\hat{A}\hat{K}_{\text{nuc}}(\mathbf{R}, \mathbf{P})\right] \tag{24}$$

and

$$\tilde{A}_{\text{nuc}}(\mathbf{R}, \mathbf{P}) = \text{Tr}_n\left[\hat{K}_{\text{nuc}}^{-1}(\mathbf{R}, \mathbf{P})\hat{A}\right] \quad . \tag{25}$$

When the Wigner function eq (14) is used, the mapping kernel and its inverse are the same, that is., $\hat{K}_{\text{nuc}}(\mathbf{x},\mathbf{p}) = \hat{K}_{\text{nuc}}^{-1}(\mathbf{x},\mathbf{p})$. The Wigner phase space function of operator $\hat{A}$ (from eq (24)) is identical to its dual (from eq (25)),

$$A_{\text{nuc}}^{W}(\mathbf{R},\mathbf{P}) = \tilde{A}_{\text{nuc}}^{W}(\mathbf{R},\mathbf{P}) \quad . \tag{26}$$

When the Husimi phase space (eq (15)) is employed, it is straightforward to show the relation between the Wigner and Husimi phase space functions (obtained from eq (24))

$$A_{\text{nuc}}^{H}(\mathbf{R},\mathbf{P}) = \exp\left[\frac{1}{4}\left(\frac{\mathrm{d}}{\mathrm{d}\mathbf{R}}\right)^{T}\boldsymbol{\Gamma}^{-1}\left(\frac{\mathrm{d}}{\mathrm{d}\mathbf{R}}\right) + \frac{\hbar^{2}}{4}\left(\frac{\mathrm{d}}{\mathrm{d}\mathbf{P}}\right)^{T}\boldsymbol{\Gamma}\left(\frac{\mathrm{d}}{\mathrm{d}\mathbf{P}}\right)\right]A_{\text{nuc}}^{W}(\mathbf{R},\mathbf{P}) \quad , \tag{27}$$

and the relation between the dual function of Husimi phase space $\tilde{A}_{\text{nuc}}^{H}(\mathbf{R},\mathbf{P})$ and the Wigner phase space function $A_{\text{nuc}}^{W}(\mathbf{R},\mathbf{P})$

$$\tilde{A}_{\text{nuc}}^{H}(\mathbf{R},\mathbf{P}) = \exp\left[-\frac{1}{4}\left(\frac{\mathrm{d}}{\mathrm{d}\mathbf{R}}\right)^{T}\boldsymbol{\Gamma}^{-1}\left(\frac{\mathrm{d}}{\mathrm{d}\mathbf{R}}\right) - \frac{\hbar^{2}}{4}\left(\frac{\mathrm{d}}{\mathrm{d}\mathbf{P}}\right)^{T}\boldsymbol{\Gamma}\left(\frac{\mathrm{d}}{\mathrm{d}\mathbf{P}}\right)\right]A_{\text{nuc}}^{W}(\mathbf{R},\mathbf{P}) \quad . \tag{28}$$

Because any choice of $f(\boldsymbol{\zeta},\boldsymbol{\eta})$ in eq (13) leads to an informationally complete representation of the continuous-variable quantum system, it is not difficult to establish the relation between different (dual) phase space functions in addition to eq (27) and eq (28).

**2.2 Mapping kernel on constraint space for discrete (electronic) degrees of freedom**

As derived first in Appendix A of Reference [134] in the spirit of Reference [133] and then in the Supporting Information of Reference [136], the kernel that maps a set of $F$ states onto constraint phase space $\mathcal{S}(\mathbf{x},\mathbf{p})$ reads

$$\hat{K}_{\text{ele}}(\mathbf{x},\mathbf{p}) = \sum_{n,m=1}^{F}\left[\frac{1}{2}\left(x^{(n)}+ip^{(n)}\right)\left(x^{(m)}-ip^{(m)}\right) - \gamma\delta_{nm}\right]|n\rangle\langle m| \tag{29}$$

and the corresponding inverse kernel is

$$\hat{K}_{\text{ele}}^{-1}(\mathbf{x},\mathbf{p}) = \sum_{n,m=1}^{F}\left[\frac{1+F}{2(1+F\gamma)^{2}}\left(x^{(n)}+ip^{(n)}\right)\left(x^{(m)}-ip^{(m)}\right) - \frac{1-\gamma}{1+F\gamma}\delta_{nm}\right]|n\rangle\langle m| \quad . \tag{30}$$

As naturally required by eq (9), constraint phase space $\mathcal{S}(\mathbf{x},\mathbf{p})$ is defined by

$$\delta\left(\sum_{n=1}^{F}\frac{\left(x^{(n)}\right)^{2}+\left(p^{(n)}\right)^{2}}{2}-(1+F\gamma)\right) \quad , \tag{31}$$

of which the area is

$$\Omega(\gamma)=\int d\mathbf{x}d\mathbf{p}\,\delta\left(\sum_{n=1}^{F}\frac{\left(x^{(n)}\right)^{2}+\left(p^{(n)}\right)^{2}}{2}-(1+F\gamma)\right) \quad . \tag{32}$$

The normalized constraint phase space is $\overline{\mathcal{S}}(\mathbf{x},\mathbf{p})=\mathcal{S}(\mathbf{x},\mathbf{p})/\Omega(\gamma)$.

Equations (29)-(32) define the mapping kernel and inverse kernel as well as constraint phase space, which are the key elements of the coordinate-momentum phase space formulation of the discrete-variable quantum system that we first established in References [133, 134] and further developed in References [57, 58, 136]. As yielded from eq (4), when the Wigner function eq (14) is used for the nuclear DOFs, the mapping Hamiltonian for the quantum Hamiltonian operator eq (1) reads

$$H_{C}(\mathbf{R},\mathbf{P};\mathbf{x},\mathbf{p};\gamma)=\frac{1}{2}\mathbf{P}^{T}\mathbf{M}^{-1}\mathbf{P}+\sum_{n,m=1}^{F}V_{mn}(\mathbf{R})\left[\frac{1}{2}\left(x^{(n)}+ip^{(n)}\right)\left(x^{(m)}-ip^{(m)}\right)-\gamma\delta_{nm}\right] . \tag{33}$$

Because $\mathbf{V}(\mathbf{R})$ is Hermitian, the mapping Hamiltonian is real. As $\mathbf{V}(\mathbf{R})$ is often a real symmetric matrix, eq (33) becomes

$$H_{C}(\mathbf{R},\mathbf{P};\mathbf{x},\mathbf{p};\gamma)=\frac{1}{2}\mathbf{P}^{T}\mathbf{M}^{-1}\mathbf{P}+\sum_{n,m=1}^{F}\left[\frac{1}{2}(x^{(n)}x^{(m)}+p^{(n)}p^{(m)})-\gamma\delta_{nm}\right]V_{mn}(\mathbf{R}) \quad , \tag{34}$$

which is the seminal Meyer-Miller Hamiltonian[155] that has extensively been implemented for nonadiabatic dynamics in the literature[4, 56, 60, 140, 156, 166-220]. In References [58, 133, 134] it is shown that there also exist other comprehensive mapping Hamiltonian models in the general coordinate-momentum phase space formulation of quantum mechanics. When the mapping Hamiltonian is employed to generate trajectory-based dynamics in the phase space formulation for a composite quantum system, we denote it the classical mapping model (CMM) approach. It satisfies the frozen nuclei limit. We use the Meyer-Miller Hamiltonian for demonstration throughout the Focus Article.

When Meyer and Miller proposed the conventional Meyer-Miller mapping Hamiltonian model for the nonadiabatic system in 1979, they did not invoke the phase space formulation. In 1997 Stock and Thoss[156] utilized the Schwinger oscillator theory of angular momentum[221, 222] to derive the Meyer-Miller mapping Hamiltonian[155]. Its LSC-IVR approximation[4] in principle includes *infinite* Wigner phase space for the finite set of (electronic) states. The applications, however, suggest that the LSC-IVR approximation in the framework of

References [4, 156, 172] is not good[174, 179, 183, 189, 197, 202, 204]. More advanced semiclassical approaches[96, 97] improve the performance but request more computational effort[174, 175]. The symmetric-window-function and other techniques have been introduced to practically overcome the drawbacks[179, 183, 189, 197, 202, 204]. Recent progress along this line is briefly summarized in Reference [140].

Equation (31) indicates that parameter $\gamma$ lies in region $(-1/F, \infty)$. It is shown that parameter $\gamma$ can be either positive or negative[133, 136] and should be interpreted as a special case of the commutator matrix[57, 58, 133, 136] rather than the conventional zero-point-energy parameter[155, 156]. There exist three key elements for a trajectory-based quantum dynamics method to evaluate the evolution of the expectation/ensemble average of a physical property, namely,

1) the EOMs of the trajectory,
2) the initial condition of the trajectory, and
3) the integral expression for the expectation/ensemble average of the physical property of interest.

In the frozen-nuclei limit, Hamilton's EOMs governed by the Meyer-Miller mapping Hamiltonian is isomorphic to exact dynamics. While it is reasonable to employ the mapping Hamiltonian to define the EOMs of the trajectory, the left two elements are also important to consider such that the trajectory-based dynamics method is consistent. The constraint coordinate-momentum phase space formulation then offers a more advanced platform to consider all the three key elements.

It is evident that eq (31) is a special choice of constraint phase space $\mathcal{S}(\mathbf{x}, \mathbf{p})$. The interpretation of parameter $\gamma$ in References [57, 58, 133, 136] hints that a more comprehensive choice of normalized constraint phase space $\bar{\mathcal{S}}(\mathbf{x}, \mathbf{p})$ is

$$\int_{-1/F}^{\infty} d\gamma \, w(\gamma) \frac{1}{\Omega(\gamma)} \delta\left( \sum_{n=1}^{F} \frac{(x^{(n)})^2 + (p^{(n)})^2}{2} - (1 + F\gamma) \right) , \qquad (35)$$

with the *quasi*-probability distribution function

$$\int_{-1/F}^{\infty} d\gamma \, w(\gamma) = 1 \quad . \qquad (36)$$

Equation (6), the integral over the mapping phase space variables for the finite discrete (electronic) DOFs then becomes

$$\int_{\overline{\mathcal{S}}(\mathbf{x},\mathbf{p})} F d\mathbf{x} d\mathbf{p}\, g(\mathbf{x},\mathbf{p})$$
$$= \int_{-1/F}^{\infty} d\gamma\, w(\gamma) \int F d\mathbf{x} d\mathbf{p} \frac{1}{\Omega(\gamma)} \delta\left(\sum_{n=1}^{F} \frac{(x^{(n)})^2 + (p^{(n)})^2}{2} - (1+F\gamma)\right) g(\mathbf{x},\mathbf{p}) \quad . \tag{37}$$

If we require that the kernel is the same as its inverse, that is,

$$\hat{K}_{\text{ele}}(\mathbf{x},\mathbf{p}) = \hat{K}_{\text{ele}}^{-1}(\mathbf{x},\mathbf{p}) = \sum_{n,m=1}^{F} \left[\frac{1}{2}(x^{(n)} + ip^{(n)})(x^{(m)} - ip^{(m)}) - \gamma \delta_{nm}\right] |n\rangle\langle m| \quad , \tag{38}$$

it is then not difficult to obtain

$$\int_{-1/F}^{\infty} d\gamma\, w(\gamma)\, \chi(\gamma) = 1 \tag{39}$$

with

$$\chi(\gamma) = F\gamma^2 + 2\gamma \quad . \tag{40}$$

(See Appendix 1 of the Supporting Information for more discussion.) Equations (35)-(40) define normalized constraint phase space $\overline{\mathcal{S}}(\mathbf{x},\mathbf{p})$, the mapping kernel and inverse kernel, and the *quasi*-probability distribution function $w(\gamma)$ of parameter $\gamma$. The weighted constraint phase space formulation for the discrete-variable quantum system is the key new theoretical result of the Focus Article. When the Wigner function eq (14) is used for the nuclear DOFs, where $\hat{K}_{\text{nuc}}(\mathbf{x},\mathbf{p}) = \hat{K}_{\text{nuc}}^{-1}(\mathbf{x},\mathbf{p})$, eq (4) is then identical to eq (5) when $\hat{A} = \hat{B}$. The mapping Hamiltonian for the quantum Hamiltonian operator eq (1) produced by either of eq (4) and eq (5) leads to the same expression as eq (34). When the mapping Hamiltonian is utilized to produce the trajectory-based dynamics for a composite system, it is denoted the weighted mapping model (wMM) approach. The frozen nuclei limit is satisfied in wMM.

Many choices are possible for the discrete or continuous version of the normalized quasi-probability distribution function $w(\gamma)$ in the weighted constraint phase space mapping theory. In the Focus Article we consider only the simplest cases of the discrete version. When but a single value of parameter $\gamma$ is chosen in eq (39), that is, $w(\gamma) = \delta(\gamma - \gamma_1)$, we obtain

$$F\gamma^2 + 2\gamma = 1 \quad , \tag{41}$$

of which the physical solution is

$$\gamma = \frac{\sqrt{1+F}-1}{F} \quad . \tag{42}$$

Equation (42) is a trivial result that was used in References [57, 114, 115, 117, 199]. In this case, the weighted constraint phase space formulation is identical to the constraint phase space formulation, and wMM becomes CMM with eq (42) when trajectory-based dynamics is considered. When only two values of parameter $\gamma$ are selected, that is,

$$w(\gamma) = \sum_{j=1}^{2} w(\gamma_j) \delta(\gamma - \gamma_j) \quad , \tag{43}$$

eq (36) and eq (39) lead to

$$\begin{aligned} w(\gamma_1) &= \frac{1 - \chi(\gamma_2)}{\chi(\gamma_1) - \chi(\gamma_2)} \\ w(\gamma_2) &= \frac{\chi(\gamma_1) - 1}{\chi(\gamma_1) - \chi(\gamma_2)} \end{aligned} \quad . \tag{44}$$

When the values of parameter $\gamma$ are close to zero or smaller than zero in region $(-1/F, \infty)$, trajectories produced by the Meyer-Miller mapping Hamiltonian eq (34) for nonadiabatic molecular dynamics are stable. For demonstration in the paper we choose

$$\gamma_1 = -\gamma_2 = \Delta \tag{45}$$

with $\Delta$ a reasonably small positive real number in region $(0, 1/F)$. Figure 3 presents the constraint coordinate-momentum phase space formulation when a single value of parameter $\gamma$ is used (Figure 3a) as well as the weighted formulation when two values of parameter $\gamma$ suggested by eq (45) are used (Figure 3b).

## 3. PHASE SPACE REPRESENTATION OF THE NONCLASSICAL FEATURE OF QUANTUM SYSTEMS

Recent advance on quantum technologies makes it possible to control and manipulate quantum states in experiment. Because the phase space formulation offers an informationally complete description of the density matrix, direct measurements of phase space of the quantum system with continuous DOFs, those of the quantum system with discrete DOFs, and those of the composite quantum system have been realized in experiment[70, 85, 223-236]. While the celebrated Wigner phase space has long been used for illustration of the negative *quasi*-probability for continuous-variable systems [227, 237], Stratonovich phase space has recently been proposed for visualization and tomography of discrete-variable systems[85, 113, 229-231, 233, 238, 239]. A combination of these two spaces has been used for illustration of nonclassical correlations or entanglement between the discrete DOF and the continuous DOF of the composite system[240, 241]. (In Appendix 3 of the Supporting Information, we briefly review Stratonovich phase space with an either SU(2) or SU($F$) structure[114, 242, 243], as well as the relationship between Stratonovich phase space and constraint coordinate-momentum phase space as already pointed out in References [57, 58].)

As coordinate-momentum phase space is well-established in classical mechanics, the formulation of (weighted) constraint coordinate-momentum phase space described in Section 2 offers a potentially useful approach for describing correlations and dynamics in the discrete-variable system as well as the composite system in quantum mechanics. When (weighted) constraint coordinate-momentum phase space is used for mapping an $F$-state system, the phase space distribution is

$$\rho_C(\mathbf{x},\mathbf{p}) = \sum_{m,n=1}^{F} \rho_{mn} K_{nm}(\mathbf{x},\mathbf{p}), \qquad (46)$$

where $\rho_{mn} = \langle m|\hat{\rho}|n\rangle$ and $K_{nm}(\mathbf{x},\mathbf{p}) = \langle n|\hat{K}_{\text{ele}}(\mathbf{x},\mathbf{p})|m\rangle$ with $\hat{K}_{\text{ele}}(\mathbf{x},\mathbf{p})$ defined in eq (38). For the sake of visualization, it is convenient to further reduce constraint phase space variables $(\mathbf{x},\mathbf{p})$ to two relevant variables, $(x^{(n)}, x^{(m)})$ or $(x^{(n)}, p^{(m)})$ for describing the correlation on arbitrary two states $|n\rangle$ and $|m\rangle$. We define the marginal function, $\mathcal{K}_{(n,m)}(x^{(n)}, x^{(m)})$, on constraint coordinate-momentum phase space (Figure 4),

$$\mathcal{K}_{(n,m)}(x^{(n)}, x^{(m)}) = \int F d\mathbf{x}_\perp d\mathbf{p} \frac{1}{\Omega(\gamma)} \delta\left(\sum_{j=1}^{F} \frac{(x^{(j)})^2 + (p^{(j)})^2}{2} - (1+F\gamma)\right) K_{nm}(\mathbf{x},\mathbf{p};\gamma), \quad (47)$$

where $\mathbf{x}_\perp$ represents all $x^{(i)}$ other than $\{x^{(n)}, x^{(m)}\}$, and that on weighted constraint phase space,

$$\mathcal{K}_{(n,m)}(x^{(n)}, x^{(m)}) = \int_{-1/F}^{\infty} d\gamma\, w(\gamma) \frac{1}{\Omega(\gamma)} \int F d\mathbf{x}_\perp d\mathbf{p}\, \delta\left(\sum_{j=1}^{F} \frac{(x^{(j)})^2 + (p^{(j)})^2}{2} - (1+F\gamma)\right) K_{nm}(\mathbf{x},\mathbf{p};\gamma).$$
(48)

Figure 5 demonstrates the case of eq (48) when the quasi-probability distribution function $w(\gamma)$ is defined by eqs (43)-(45) where two symmetrical values of parameter $\gamma$ are used. Similar definitions are also applied for $\mathcal{K}_{(n,m)}(x^{(n)}, p^{(m)})$. The explicit formula of these marginal functions can be derived by using the integral techniques (where we use Wick's theorem[244, 245]) in Appendix 1 of the Supporting Information.

Figures 4-5 demonstrate a composite system that consists of a discrete DOF for spin-1/2 and a continuous DOF for a harmonic oscillator. The marginal joint distribution function of the composite system reads

$$\rho_C^{(n,m)}(\mathbf{R},\mathbf{P};x^{(n)},x^{(m)})=\mathrm{Tr}_{n,e}\left[\hat{\rho}\hat{K}_{nuc}(\mathbf{R},\mathbf{P})\otimes|n\rangle\langle m|\mathcal{K}_{(n,m)}(x^{(n)},x^{(m)})\right]. \tag{49}$$

The marginal quasi-probability distribution functions of the continuous variable for both the pure state and the mixed state are presented in Figure 4a, where infinite Wigner phase space is employed. The marginal functions of the discrete variables (based on eq (47)) of the spin-1/2 system read

$$\begin{pmatrix} \mathcal{K}_{\uparrow\uparrow}(x^{(1)},x^{(2)}) & \mathcal{K}_{\uparrow\downarrow}(x^{(1)},x^{(2)}) \\ \mathcal{K}_{\downarrow\uparrow}(x^{(1)},x^{(2)}) & \mathcal{K}_{\downarrow\downarrow}(x^{(1)},x^{(2)}) \end{pmatrix}$$
$$= \frac{1}{2\pi(1+2\gamma)} \begin{pmatrix} 1+\frac{1}{2}(x^{(1)})^2 - \frac{1}{2}(x^{(2)})^2 & x^{(1)}x^{(2)} \\ x^{(1)}x^{(2)} & 1-\frac{1}{2}(x^{(1)})^2 + \frac{1}{2}(x^{(2)})^2 \end{pmatrix}, \tag{50}$$

where notations $\uparrow,\downarrow$ are used to represent the two discrete states.

The marginal functions for the discrete variable are demonstrated on contraint coordinate-momentum phase space in Figure 4b and on weighted constraint space in Figure 5b. More interestingly, the identical angular behaviour and the radial cancellation behaviour of two weighted components lead to a hollow ring structure on weighted constraint phase space (Figure 5a, also see Appendix 4 of the Supporting Information). The difference between the Schrodinger cat state and the mixed state is distinct in either Figure 4b on constraint space or Figure 5b on weighted constraint space.

The marginal joint function of a pure Bell entangled state, $(|0\rangle|\downarrow\rangle + |1\rangle|\uparrow\rangle)/2$, of the composite system is demonstrated in Figure 4c (by adopting the similar strategy of References [240, 241]), where constraint coordinate-momentum phase space is used for the discrete DOF at each grid, as well as in Figure 5c where weighted constraint space is employed for the discrete DOF at each grid. The two-dimensional grids represent variables $(R,P)$ of infinite Wigner phase space for the continuous DOF in either of Figure 4c and Figure 5c. When the pure Bell entangled state is studied, both Figure 4c and Figure 5c clearly demonstrate a Gaussian decay of the joint marginal function against Wigner phase space variables $(R,P)$ of the continuous DOF. Either Figure 4c or Figure 5c also shows the pattern of the correlation between the continuous DOF and the discrete DOF. It is convenient to distinguish the pure Bell entangled state, $(|0\rangle|\downarrow\rangle + |1\rangle|\uparrow\rangle)/2$, from the direct product of the Schrodinger cat states, $(|0\rangle+|1\rangle)\otimes(|\uparrow\rangle+|\downarrow\rangle)/2$, when the hybrid representation of the general coordinate-momentum phase space is used.

## 4. DYNAMICS OF COMPOSITE QUANTUM SYSTEMS

The quantum Liouville theorem can be expressed as a generalized Moyal bracket on hybrid coordinate-momentum phase space. When the Poisson bracket for classical Hamilton's EOMs governed by the mapping Hamiltonian, eq (34), is used to approximate the generalized Moyal bracket on phase space[57, 58], we have CMM when constraint space is used, and wMM when weighted constraint space is employed. . (Please see Appendices 2, 3, and 5 of the Supporting Information for more discussion.) We compare the new wMM and CMM approaches to Ehrenfest dynamics[246, 247] as well as the fewest-switches surface hopping (FSSH) method[248-250], two prevailing trajectory-based dynamics methods for a few typical composite quantum systems. (In this section we set $\hbar = 1$ for simplicity if it is not specifically stated).

### 4.1 Equations of motion governed by the mapping Hamiltonian

In eq (1), the 'complete' set of diabatic states $\{|n\rangle\}$ is independent of nuclear coordinate/configuration $\mathbf{R}$. The mapping variables for discrete (electronic) DOFs, $(\mathbf{x}, \mathbf{p})$, are independent of $\mathbf{R}$. Define $\mathbf{g} = \mathbf{x} + i\mathbf{p}$. The EOMs governed by eq (33), the mapping Hamiltonian of eq (1), then read,

$$\dot{\mathbf{g}} = -i\mathbf{V}(\mathbf{R})\mathbf{g} \quad . \tag{51}$$

$$\dot{\mathbf{R}} = \mathbf{M}^{-1}\mathbf{P} \tag{52}$$

$$\dot{\mathbf{P}} = -\sum_{n,m=1}^{F} \left(\nabla_{\mathbf{R}} V_{mn}(\mathbf{R})\right)\left(\frac{1}{2}\left(x^{(n)} + ip^{(n)}\right)\left(x^{(m)} - ip^{(m)}\right) - \gamma\delta_{nm}\right) \quad . \tag{53}$$

Diabatic potential matrix $\mathbf{V}(\mathbf{R})$ is Hermitian, so is the force matrix, $\{\nabla_{\mathbf{R}} V_{mn}(\mathbf{R})\}$. It is trivial to verify that the mean force of the right-hand side (RHS) of eq (53) is always real. When $\mathbf{V}(\mathbf{R})$ is a real symmetric matrix, the EOMs become

$$\begin{aligned}
\dot{\mathbf{x}} &= \mathbf{V}(\mathbf{R})\mathbf{p} \\
\dot{\mathbf{p}} &= -\mathbf{V}(\mathbf{R})\mathbf{x} \\
\dot{\mathbf{R}} &= \mathbf{M}^{-1}\mathbf{P} \\
\dot{\mathbf{P}} &= -\sum_{n,m=1}^{F} \left(\nabla_{\mathbf{R}} V_{mn}(\mathbf{R})\right)\left[\frac{1}{2}(x^{(n)}x^{(m)} + p^{(n)}p^{(m)}) - \gamma\delta_{nm}\right]
\end{aligned} \tag{54}$$

Consider the full Hamiltonian of nuclei and electrons of the molecular system,

$$\hat{H} = \frac{1}{2}\hat{\mathbf{P}}^T \mathbf{M}^{-1}\hat{\mathbf{P}} + \hat{H}_{el}(\hat{\mathbf{R}}) \quad , \tag{55}$$

where $\hat{H}_{el}(\hat{\mathbf{R}})$ is the electronic Hamiltonian. Its representation in the diabatic basis reads

$$\hat{H}_{el}(\mathbf{R}) = \sum_{n,m} V_{nm}(\mathbf{R}) |n\rangle\langle m| \tag{56}$$

and that in the adiabatic basis is

$$\hat{H}_{el}(\mathbf{R}) = \sum_{k} E_k(\mathbf{R}) |\phi_k(\mathbf{R})\rangle\langle\phi_k(\mathbf{R})| \quad , \tag{57}$$

where $E_k(\mathbf{R})$ denotes the adiabatic potential energy surface of the $k$-th adiabatic electronic state. Assume that the unitary transformation between a set of diabatic basis states, $\{|m\rangle\}$, and a set of adiabatic basis states, $\{|\phi_k(\mathbf{R})\rangle\}$, is

$$\begin{aligned} |\phi_k(\mathbf{R})\rangle &= \sum_{m} U_{mk}(\mathbf{R}) |m\rangle \\ |n\rangle &= \sum_{k} U_{nk}^*(\mathbf{R}) |\phi_k(\mathbf{R})\rangle \end{aligned} \quad , \tag{58}$$

where $U_{mk}(\mathbf{R}) = \langle m|\phi_k(\mathbf{R})\rangle$. This states the diagonalization of the diabatic potential matrix,

$$\sum_{n,m} U_{nj}^*(\mathbf{R}) V_{nm}(\mathbf{R}) U_{mk}(\mathbf{R}) = E_k(\mathbf{R}) \delta_{kj} \quad , \tag{59}$$

or equivalently,

$$V_{mn}(\mathbf{R}) = \sum_{k} U_{mk}(\mathbf{R}) E_k(\mathbf{R}) U_{nk}^*(\mathbf{R}) \quad . \tag{60}$$

Define the nonadiabatic coupling vector,

$$\mathbf{d}_{mn}(\mathbf{R}) = \left\langle \phi_m(\mathbf{R}) \left| \frac{\partial \phi_n(\mathbf{R})}{\partial \mathbf{R}} \right. \right\rangle \quad . \tag{61}$$

It is trivial to show

$$\mathbf{d}_{mn}(\mathbf{R}) = -\mathbf{d}_{nm}^*(\mathbf{R}) \tag{62}$$

because of the orthonormality of the basis set, that is, $\langle \phi_m(\mathbf{R})|\phi_n(\mathbf{R})\rangle = \delta_{mn}$. We then obtain

$$\begin{aligned} \nabla_{\mathbf{R}} U_{mk}^*(\mathbf{R}) &= \langle \nabla_{\mathbf{R}} \phi_k(\mathbf{R})|m\rangle = \sum_{n} \langle \nabla_{\mathbf{R}} \phi_k(\mathbf{R})|\phi_n\rangle\langle\phi_n|m\rangle \\ &= \sum_{n} \mathbf{d}_{nk}^*(\mathbf{R}) U_{mn}^*(\mathbf{R}) = -\sum_{n} \mathbf{d}_{kn}(\mathbf{R}) U_{mn}^*(\mathbf{R}) \end{aligned} \tag{63}$$

and

$$\nabla_{\mathbf{R}} U_{mk}(\mathbf{R}) = -\sum_{n} \mathbf{d}^*_{kn}(\mathbf{R}) U_{mn}(\mathbf{R}) = \sum_{n} U_{mn}(\mathbf{R}) \mathbf{d}_{nk}(\mathbf{R}) \quad . \tag{64}$$

Below we show the explicit form of the EOMs, eqs (51)-(53), under the diabatic-to-adiabatic transformation, eq (58).

The covariant transformation for mapping variables corresponding to the diabatic-to-adiabatic transformation, eq (58), reads

$$\tilde{x}^{(n)}(\mathbf{R}) + i\tilde{p}^{(n)}(\mathbf{R}) = \sum_{m} U^*_{mn}(\mathbf{R}) \left( x^{(m)} + ip^{(m)} \right) \tag{65}$$

or

$$x^{(n)} + ip^{(n)} = \sum_{m} U_{nm}(\mathbf{R}) \left( \tilde{x}^{(m)}(\mathbf{R}) + i\tilde{p}^{(m)}(\mathbf{R}) \right) \quad . \tag{66}$$

Denote $\tilde{\mathbf{g}}(\mathbf{R}) = \tilde{\mathbf{x}}(\mathbf{R}) + i\tilde{\mathbf{p}}(\mathbf{R})$. Equations (65)-(66) become

$$\begin{aligned} \tilde{\mathbf{g}}(\mathbf{R}) &= \mathbf{U}^\dagger(\mathbf{R}) \mathbf{g} \\ \mathbf{g} &= \mathbf{U}(\mathbf{R}) \tilde{\mathbf{g}}(\mathbf{R}) \end{aligned} \quad . \tag{67}$$

The electronic mapping kernel, eq (29), is

$$\hat{K}_{\text{ele}} = \sum_{n,m} \left[ \frac{1}{2} \left( \tilde{x}^{(n)} + i\tilde{p}^{(n)} \right) \left( \tilde{x}^{(m)} - i\tilde{p}^{(m)} \right) - \gamma \delta_{nm} \right] |\phi_n\rangle \langle \phi_m| \quad , \tag{68}$$

under the transformation for a specific nuclear configuration, $\mathbf{R}$. Substitution of eq (63) into eq (65) yields

$$\nabla_{\mathbf{R}} \left( \tilde{x}^{(n)}(\mathbf{R}) + i\tilde{p}^{(n)}(\mathbf{R}) \right) = -\sum_{k} \mathbf{d}_{nk}(\mathbf{R}) \left( \tilde{x}^{(k)}(\mathbf{R}) + i\tilde{p}^{(k)}(\mathbf{R}) \right) \quad . \tag{69}$$

The total time derivative of $\tilde{x}^{(n)} + i\tilde{p}^{(n)}$ reads

$$\begin{aligned} \frac{\mathrm{d}}{\mathrm{d}t} \left( \tilde{x}^{(n)} + i\tilde{p}^{(n)} \right) &= \sum_{m} U^*_{mn}(\mathbf{R}) \left( \frac{\mathrm{d}}{\mathrm{d}t} (x^{(m)} + ip^{(m)}) \right) + \sum_{m} \left( \frac{\mathrm{d}}{\mathrm{d}t} U^*_{mn}(\mathbf{R}) \right) (x^{(m)} + ip^{(m)}) \\ &= -i \sum_{k} \delta_{nk} E_k(\mathbf{R}) (\tilde{x}^{(k)} + i\tilde{p}^{(k)}) - \sum_{k} \dot{\mathbf{R}} \cdot \mathbf{d}_{nk}(\mathbf{R}) (\tilde{x}^{(k)} + i\tilde{p}^{(k)}) \\ &= -i \sum_{k} \left[ E_k(\mathbf{R}) \delta_{nk} - i\dot{\mathbf{R}} \cdot \mathbf{d}_{nk}(\mathbf{R}) \right] (\tilde{x}^{(k)} + i\tilde{p}^{(k)}) \end{aligned} \tag{70}$$

Equation (70) is the EOMs for mapping variables of electronic DOFs in the adiabatic representation.

We then consider the EOMs of nuclear mapping variables under the transformation eq (58). Equation (52) remains invariant under the transformation. Substitution of eqs (60), (63), (64), and 69 into eq (53) produces

$$\begin{aligned}
\dot{\mathbf{P}} &= -\sum_{n,m=1}^{F} (\nabla_\mathbf{R} V_{mn}(\mathbf{R})) \left[ \frac{1}{2}(x^{(n)} + ip^{(n)})(x^{(m)} - ip^{(m)}) - \gamma \delta_{nm} \right] \\
&= -\sum_{n,m} \nabla_\mathbf{R} \left( \sum_k U_{mk}(\mathbf{R}) E_k(\mathbf{R}) U^*_{nk}(\mathbf{R}) \right) \left[ \frac{1}{2}(x^{(n)} + ip^{(n)})(x^{(m)} - ip^{(m)}) - \gamma \delta_{nm} \right] \\
&= \sum_{k,l} \mathbf{d}_{lk}(\mathbf{R})(E_l(\mathbf{R}) - E_k(\mathbf{R})) \left[ \frac{1}{2}(\tilde{x}^{(k)} + i\tilde{p}^{(k)})(\tilde{x}^{(l)} - i\tilde{p}^{(l)}) - \gamma \delta_{kl} \right] \\
&\quad - \sum_{k,l} \nabla_\mathbf{R} E_k(\mathbf{R}) \delta_{kl} \left[ \frac{1}{2}(\tilde{x}^{(k)} + i\tilde{p}^{(k)})(\tilde{x}^{(l)} - i\tilde{p}^{(l)}) - \gamma \delta_{kl} \right]
\end{aligned} \quad (71)$$

Since force matrix $\{\mathbf{F}_{kl} = \nabla_\mathbf{R} E_k(\mathbf{R}) \delta_{kl} + (E_k(\mathbf{R}) - E_l(\mathbf{R})) \mathbf{d}_{lk}(\mathbf{R})\}$ is Hermitian, the mean force of the RHS of eq (71) stays real. Under the diabatic-to-adiabatic transformation, eq (58), the EOMs of nuclear phase variables (eqs (52)-(53)) are then recast into

$$\begin{aligned}
\dot{\mathbf{R}} &= \mathbf{M}^{-1} \mathbf{P} \\
\dot{\mathbf{P}} &= -\sum_{k,l} \left[ \nabla_\mathbf{R} E_k(\mathbf{R}) \delta_{kl} + (E_k(\mathbf{R}) - E_l(\mathbf{R})) \mathbf{d}_{lk}(\mathbf{R}) \right] \left[ \frac{1}{2}(\tilde{x}^{(k)} + i\tilde{p}^{(k)})(\tilde{x}^{(l)} - i\tilde{p}^{(l)}) - \gamma \delta_{kl} \right]
\end{aligned} \quad (72)$$

Define the effective potential matrix, $\mathbf{V}^{(\text{eff})}$, whose element is a function of the nuclear phase variables,

$$V^{(\text{eff})}_{nk}(\mathbf{R}, \mathbf{P}) = E_n(\mathbf{R}) \delta_{nk} - i\dot{\mathbf{R}} \cdot \mathbf{d}_{nk}(\mathbf{R}) = E_n(\mathbf{R}) \delta_{nk} - i\mathbf{M}^{-1} \mathbf{P} \cdot \mathbf{d}_{nk}(\mathbf{R}) \quad . \quad (73)$$

A more compact form of eq (70) for the electronic phase variables becomes

$$\dot{\tilde{\mathbf{g}}} = -i \mathbf{V}^{(\text{eff})}(\mathbf{R}, \mathbf{P}) \tilde{\mathbf{g}} \quad . \quad (74)$$

Equations (72) and (74) are the final EOMs under the covariant transformation eq (65).

When the electronic wavefunction of the basis set is always real, that is, $\langle r | \phi_n(\mathbf{R}) \rangle$ is real for any $n$, which is often the case for molecular systems, eq (62) leads to

$$\mathbf{d}_{mn}(\mathbf{R}) = -\mathbf{d}_{nm}(\mathbf{R}) \quad . \quad (75)$$

Equation (72) is simplified to

$$\begin{aligned}
\dot{\mathbf{R}} &= \mathbf{M}^{-1} \mathbf{P} \\
\dot{\mathbf{P}} &= -\sum_{k,l} \left[ \nabla_\mathbf{R} E_k(\mathbf{R}) \delta_{kl} + (E_k(\mathbf{R}) - E_l(\mathbf{R})) \mathbf{d}_{lk}(\mathbf{R}) \right] \left[ \frac{1}{2}(\tilde{x}^{(k)} \tilde{x}^{(l)} + \tilde{p}^{(k)} \tilde{p}^{(l)}) - \gamma \delta_{kl} \right]
\end{aligned} \quad (76)$$

Note that the mapping Hamiltonian of eq (33) (obtained in the diabatic representation) becomes

$$H_C(\mathbf{R}, \mathbf{P}, \mathbf{x}(\tilde{\mathbf{x}}, \tilde{\mathbf{p}}), \mathbf{p}(\tilde{\mathbf{x}}, \tilde{\mathbf{p}})) = \frac{1}{2} \mathbf{P}^T \mathbf{M}^{-1} \mathbf{P} + \sum_{n=1}^{F} E_n(\mathbf{R}) \left( \frac{1}{2} \left( (\tilde{x}^{(n)}(\mathbf{R}))^2 + (\tilde{p}^{(n)}(\mathbf{R}))^2 \right) - \gamma \right) \quad (77)$$

under the transformation defined by eq (60) and eq (65). The new EOMs, eq (72) and eq (74), conserve the mapping Hamiltonian of eq (77). The diabatic-to-adiabatic transformation depends on nuclear coordinate $\mathbf{R}$, which is also a time-dependent variable of the evolution. The time-dependent canonical transformation for the Hamiltonian system yields a new set of EOMs by the chain rule[251].

In eqs (71)-(74) and eqs (76)-(77) $\mathbf{P}$ corresponds to the mapping momentum in the diabatic representation, but *not* the canonical momentum in the adiabatic representation because eq (71) is *not* generated from Hamilton's equations of motion. Equations (74) and (76) share a similar form to the EOMs proposed by Cotton *et al* [186] and discussed in the Supporting Information of Reference [57]. Define the covariant transformation for nuclear phase variables,

$$\tilde{\mathbf{R}} = \mathbf{R}$$
$$\tilde{\mathbf{P}} = \mathbf{P} + i\sum_{m,n}\left[\frac{1}{2}\left(\tilde{x}^{(n)} + i\tilde{p}^{(n)}\right)\left(\tilde{x}^{(m)} - i\tilde{p}^{(m)}\right) - \gamma\delta_{nm}\right]\mathbf{d}_{mn}(\mathbf{R}) \quad . \tag{78}$$

The Hamiltonian of eq (77) becomes

$$H_C\left(\tilde{\mathbf{R}}, \tilde{\mathbf{P}}, \tilde{\mathbf{x}}, \tilde{\mathbf{p}}\right) = \frac{1}{2}\mathbf{P}(\tilde{\mathbf{P}}, \tilde{\mathbf{x}}, \tilde{\mathbf{p}}, \tilde{\mathbf{R}})^T \mathbf{M}^{-1}\mathbf{P}(\tilde{\mathbf{P}}, \tilde{\mathbf{x}}, \tilde{\mathbf{p}}, \tilde{\mathbf{R}})$$
$$+ \sum_{n=1}^{F} E_n(\tilde{\mathbf{R}})\left(\frac{1}{2}\left(\left(\tilde{x}^{(n)}(\tilde{\mathbf{R}})\right)^2 + \left(\tilde{p}^{(n)}(\tilde{\mathbf{R}})\right)^2\right) - \gamma\right) \quad , \tag{79}$$

of which the canonical variables are $\{\tilde{\mathbf{R}}, \tilde{\mathbf{P}}, \tilde{\mathbf{x}}, \tilde{\mathbf{p}}\}$ instead of $\{\mathbf{R}, \mathbf{P}, \tilde{\mathbf{x}}, \tilde{\mathbf{p}}\}$. (See more discussions in Appendix 2 of the Supporting Information). The mapping diabatic momentum, $\mathbf{P}$, is related to the kinematic momentum of the adiabatic representation. Although we can directly use Hamilton's EOMs for $\{\tilde{\mathbf{R}}, \tilde{\mathbf{P}}, \tilde{\mathbf{x}}, \tilde{\mathbf{p}}\}$, it is more convenient to employ the EOMs for $\{\mathbf{R}, \mathbf{P}, \tilde{\mathbf{x}}, \tilde{\mathbf{p}}\}$ instead to avoid the derivative of nonadiabatic coupling terms. This is indeed the strategy suggested by Cotton *et al* [186]. When the initial condition does not involve nonadiabatic coupling terms, the sampling of $\mathbf{P}$ in the diabatic representation is the same for that of $\tilde{\mathbf{P}}$ in the adiabatic representation. This is the case in the following applications, where FSSH has to be used in the adiabatic representation. By applying the covariance relation under the diabatic-to-adiabatic transformation, the EOMs on mapping phase space are independent of the representation of the (electronic) basis set, which is also the merit of Ehrenfest dynamics.

We note that either eq (51) or eq (74) can analytically be solved by a symplectic approach that employs an exact propagator on electronic phase space at each nuclear phase point. For example, for eq (74) we use

$$\tilde{\mathbf{U}}(\mathbf{R},\mathbf{P};\Delta t)=\exp[-i\Delta t\mathbf{V}^{(\text{eff})}] \qquad (80)$$

such that the evolution of electronic phase variables follows $\tilde{\mathbf{g}}(t+\Delta t)=\tilde{\mathbf{U}}(\mathbf{R},\mathbf{P};\Delta t)\tilde{\mathbf{g}}(t)$.

We then test a range of benchmark systems, including two-site dissipative models, Tully's scattering models, atomic systems in cavity interacted with a number of field modes, and linear vibronic coupling model systems that involve the conical intersection[137, 252-254]. They are typical composite quantum systems in chemistry, physics, condensed matter science, quantum optics, and quantum information.

### 4.2 Spin-boson models at low-temperature in condensed phase

The first model illustrated is the spin-boson model, which describes a two-site system interacted with an environmental bath in condense phase. It is also a simplified model for electron transfer and energy transfer in chemical and biological reactions. Several numerically exact benchmark methods for solving the spin-boson model include quasi-adiabatic propagator path integral (QuAPI)[255-258] and more efficient small matrix PI (SMatPI)[259, 260], hierarchy equations of motion (HEOM)[261-269], and (multi-layer) multi-configuration time-dependent Hartree [(ML-)MCTDH] [270-276]. Quantum dynamics of the spin-boson model exhibits interesting dissipative characters, of which the asymptotic behaviours are often missed by either of Ehrenfest dynamics and FSSH in the low temperature regime[58]. Spin-boson models with strong coupling in the low temperature regime presents challenging tests for trajectory-based dynamics methods.

The Hamiltonian of the spin-boson model is divided to three parts, $\hat{H}=\hat{H}_s+\hat{H}_b+\hat{H}_{sb}$. Here $\hat{H}_s=\varepsilon\hat{\sigma}_z+\Delta_c\hat{\sigma}_x$ describes a two-site system with the bias, $\varepsilon$, and tunneling $\Delta_c$, while the bath part of the Hamiltonian is discretized into a combination of a number of quantum harmonic oscillators $\hat{H}_b=\sum_{j=1}^{N_b}\left(\hat{P}_j^2+\omega_j^2\hat{R}_j^2\right)/2$. The system-bath coupling adopts a bilinear interaction, $\hat{H}_{sb}=-\sum_{j=1}^{N_b}c_j\hat{R}_j\hat{\sigma}_z$. Here we use an Ohmic bath spectral density $J(\omega)=(\pi/2)\alpha\omega e^{-\omega/\omega_c}$, where $\alpha$ is the Kondo parameter and $\omega_c$ is the cut-off frequency. Its discrete frequencies and coupling strengths $\{\omega_j,c_j\}$ are sampled[277-279] from

$$\begin{cases}\omega_j=-\omega_c\ln\left[1-j/(1+N_b)\right]\\ c_j=\omega_j\sqrt{\alpha\omega_c/(1+N_b)}\end{cases},\quad j=1,\cdots,N_b \qquad (81)$$

The initial density is set as $|1\rangle_s \langle 1|_s \otimes \hat{\rho}_b$, where the system is in excited state $|1\rangle_s$ while all bath modes are at thermal equilibrium with $\hat{\rho}_b = e^{-\beta \hat{H}_b}/Z_b$. Initial nuclear DOFs are sampled from the Wigner distribution of $\hat{\rho}_b$, while initial electronic DOFs are sampled from (weighted) constraint coordinate-momentum phase space $\mathcal{S}(\mathbf{x},\mathbf{p})$. The continuous spectral density is discretized into $N_b = 300$ effective bath modes to guarantee numerical convergence in simulations.

In Figure 6, we demonstrate results produced by wMM with parameter $\Delta = 0.05$, by wMM with $\Delta = 0.1$, and by CMM with $\gamma = (\sqrt{F+1}-1)/F = 0.366$ that is a special case of CMM of Reference [136]. Numerically exact results, as well as results yielded by Ehrenfest dynamics and FSSH, are also shown for comparison. Figure 6 indicates that wMM, as well as CMM, outperforms both Ehrenfest dynamics and FSSH dynamics, either for short-time coherences or for long-time dissipations.

**4.3 Tully's gas phase scattering models**

Tully's scattering models[248] mimic different intersection types of molecular systems, which have widely been tested for various nonadiabatic dynamics methods. They describe a two-state Hamiltonian with a central coupling area and asymptotic plateau regions where diabatic potential function $V_{nn}(R \to \pm\infty)$ is flat. All the three models, including the single avoided crossing (SAC), dual avoided crossing (DAC), and extended coupling region (ECR) problems, are used in our numerical tests.

Atomic units are used in the simulations of the Tully models. The SAC model (Panel a1 of Figure 7) describes the simplest but essential surface crossing in molecular systems. In the diabatic representation, its diagonal potential energy surfaces (PESes) are $V_{11} = -V_{22} = A(1-e^{-B|R|})\text{sgn}(R)$ and off-diagonal coupling terms are $V_{12} = V_{21} = Ce^{-DR^2}$. Here, the parameters are $A = 0.01$, $B = 1.6$, $C = 0.005$, and $D = 1.0$. The DAC model (Panel b1 of Figure 7) includes two crossing points, thus different (electronic) paths are interfered with the dependence on the initial momentum. Its diagonal PESes are $V_{11} = 0$ and $V_{22} = -Ae^{-BR^2} + E_0$, and off-diagonal coupling terms are $V_{12} = V_{21} = Ce^{-DR^2}$ in the diabatic representation with parameters $A = 0.10$, $B = 0.28$, $E_0 = 0.05$, $C = 0.015$ and $D = 0.06$. The ECR model in the diabatic representation (Panel c1 of Figure 7) has diagonal PESes $V_{11} = -V_{22} = E_0$ and coupling terms $V_{12} = V_{21} = C\left[e^{BR}\Theta(-R) + (2-e^{-BR})\Theta(R)\right]$, with $E_0 = -0.0006$, $B = 0.9$, $C = 0.1$. Here

$\Theta(R)$ is the Heaviside function of coordinate $R$. The adiabatic PESes and nonadiabatic coupling vector of the ECR model are also illustrated in Panel c2 of Figure 7.

We investigate the transmission and reflection coefficients of each state. In the simulations, the initial condition is a nuclear wavepacket, $\Psi(R;t=0) \propto \exp[-\alpha(R-R_0)^2/2 + i(R-R_0)P_0]$ (here we adopt $\hbar = 1$), occupied in state $1$, where $\alpha = 1$ is the Gaussian width parameter, and $R_0$ and $P_0$ are the initial average coordinate and momentum. The initial average coordinate is set at $R_0 = -3.8$, $-10$, and $-13$ for the SAC, DAC and ECR models, respectively. The initial Wigner distribution for the nuclear DOF is then $\rho_W^{nuc}(R,P) \propto \exp[-\alpha(R-R_0)^2 - (P-P_0)^2/\alpha]$.

Figure 7a shows that all methods are capable of quantitatively describing transmission coefficients in (diabatic) state 1 and state 2 of the SAC model. Figure 7b demonstrates that either wMM or CMM outperforms Ehrenfest dynamics and FSSH in predicting the peak shape when the initial momentum is relatively high, for example, $P_0 \geq 15$ au. This indicates that the trajectory-based approximate dynamics approaches in the mapping phase space formulation are good for fast processes in the gas phase composite/nonadiabatic system. However, the performance of either wMM or CMM in the low initial momentum region should be improved. It is important to note that the EOMs of wMM/CMM are invariant with the representation of the electronic state, as described in the Supporting Information of Reference [57]. (More discussion is also available in Appendix 2 of the Supporting Information.) That is, both the diabatic and adiabatic representations produce the same results for wMM or CMM, which is often not satisfied in FSSH and other nonadiabatic dynamics approaches.

For the ECR model of Figure 7c, the numerically exact DVR solution indicates an energy threshold for a bifurcation. Ehrenfest dynamics totally misses the step-like behaviours for the transmission coefficient in state 1, and for the reflection coefficient in either state 1 or state 2. CMM greatly improves over Ehrenfest dynamics. It is more encouraging that wMM is capable of faithfully describing such step-like behaviours. Tully's original FSSH algorithm is not able to well describe the ECR model[248], but a modified version for treating frustrated hopping of FSSH (e.g., see Reference [250]) is capable of qualitatively capturing the step-like behaviours. As shown in Figure 7c, in comparison to the traditional FSSH approach[248, 250], the overall performance of wMM for the ECR model is better.

### 4.4 Atom/molecule-in-cavity models of quantum electrodynamic light-matter systems

The cavity quantum electrodynamics (cQED) focuses on studying the interaction between light and a multi-level system (e.g., an atom or a molecule) in an optical cavity, which has many applications in the field of quantum information and quantum computation. There exist many interesting and important phenomena in cQED, for example, the Purcell effect when the coupling is weak and the vacuum Rabi splitting when the coupling becomes strong[280-294]. When the general atomic/molecular system is coupled to multi-cavity modes, it is often intractable to solve the exact evolution in real time due to the curse of dimensionality. We test wMM for two typical models that describe an imprisoned multi-level atom coupled with a series of optical modes in a one-dimensional lossless cavity[57, 206, 295-298].

The total Hamiltonian consists of three parts. The optical field is depicted by $N$ effective modes

$$\hat{H}_p = \sum_{j=1}^{N} \frac{1}{2}\left(\hat{P}_j^2 + \omega_j^2 \hat{R}_j^2\right) \quad , \tag{82}$$

where $\{\hat{R}_j, \hat{P}_j\}$ denote the canonical coordinate-momentum variables of $j$-th optical field mode with the corresponding photonic frequency $\omega_j$. The atomic system is described by $\hat{H}_a = \sum_{n=1}^{F} \varepsilon_n |n\rangle\langle n|$ with $\varepsilon_n$ representing the $n$-th atomic energy level. Employing the dipole approximation, one can formulate the interaction between atom and optical field as

$$\hat{H}_c = \sum_{n \neq m}^{F} \left( \sum_{j=1}^{N} \omega_j \lambda_j(r_0) \hat{R}_j \right) \mu_{nm} |n\rangle\langle m| \quad . \tag{83}$$

Here $\mu_{nm}$ denotes the transitional dipole moment between the $n$-th and $m$-th atomic levels, and the coupling between the $j$-th mode and the atom is

$$\lambda_j(r_0) = \sqrt{\frac{2}{\varepsilon_0 L}} \sin\left(\frac{j\pi r_0}{L}\right) \quad , \tag{84}$$

where $L$ is the volume length of cavity, $\varepsilon_0$ denotes the vacuum permittivity, and $r_0$ represents the location of the atom. In the simulation, the volume length of the cavity is set to 236200 au and the atom is frozen at the central location, that is, $r_0 = L/2$. The optical field is depicted by 400 standing-wave modes in cavity, of which the $j$-th frequency is $\omega_j = j\pi c/L$ with $c$ the light speed in vacuum. We use two benchmark models for studying cQED processes, a three-level model with $\varepsilon_1 = -0.6738$, $\varepsilon_2 = -0.2798$, $\varepsilon_3 = -0.1547$, $\mu_{12} = -1.034$, $\mu_{23} = -2.536$ (all in atomic units), and a reduced two-level model where only the two lowest atomic levels are employed.

The highest atomic level of each model is initially occupied with no photon in cavity, that is, all cavity modes are in the corresponding vacuum state. The spontaneous emission occurs at the beginning, the released photon evolves in the cavity, and the re-absorption and re-emission happen later when the photon is reflected to meet the atom. Figure 8 shows the population transfer of each atomic level of the two models. The wMM results are compared with CMM, Ehrenfest dynamics, FSSH, and exact results[296, 297]. Results of Ehrenfest dynamics and of FSSH significantly deviate from exact results even since very short time, while CMM and wMM yield much more reasonable descriptions for all energy levels, including the transfer behaviour at short time and the revival at around $t = 1800$ au. The wMM approach shows overall better performance than CMM in most of the cases. Figure 8 implies that the trajectory-based methods in the general coordinate-momentum phase space formulation will be useful for studying cQED phenomena in the field of quantum optics and quantum information.

### 4.5 Linear vibronic coupling model for the molecular system involving the conical intersection

The conical intersection widely exists in molecular systems and plays a central role in many photophysical and photochemical phenomena[137, 141, 216, 252, 253, 299-302]. The linear vibronic coupling model (LVCM) is the simplest but effective model widely used to describe dynamic properties around the conical intersection region, of which Hamiltonian in the diabatic representation is

$$\hat{H} = \hat{H}_0 + \hat{H}_l + \hat{H}_c \quad . \tag{85}$$

Here, $\hat{H}_0 = \sum_{k=1}^{N} \omega_k \left( \hat{P}_k^2 + \hat{R}_k^2 \right)/2$ is the zeroth-order harmonic oscillator Hamiltonian in normal-mode space of the electronic ground state, where $\hat{P}_k, \hat{R}_k \, (k = 1, \cdots, N)$ denote the $k$-th effective weighted normal-mode variables with frequency $\omega_k$ (i.e., $P_k = p_k / \sqrt{\omega_k}$, $R_k = \sqrt{\omega_k} r_k$, where $p_k, r_k$ are the canonical momentum, and canonical coordinate of $k$-th normal-mode). In eq (85), $\hat{H}_l = \sum_{n=1}^{F} \left( E_n + \sum_{k=1}^{N} \kappa_k^{(n)} \hat{R}_k \right) |n\rangle\langle n|$ contains the vertical excitation energy, $E_n \, (n = 1, \cdots, F)$ of $F$ electronic states, and the linear coupling term $\kappa_k^{(n)}$ of each nuclear DOF for diagonal Hamiltonian elements, while $\hat{H}_c = \sum_{n \neq m}^{F} \left( \sum_{k=1}^{N} \lambda_k^{(nm)} \hat{R}_k \right) |n\rangle\langle m|$ includes linear coupling $\lambda_k^{(nm)}$ for each normal-mode between two different electronic states, $|n\rangle$ and $|m\rangle$.

A typical two-level 3-mode LVCM describes the S1/S2 conical intersection of the pyrazine molecule. The parameters of this model are fitted from semi-empirical electronic structure calculations by Schneiders and

Domcke in Reference [303]. The excitation energies for the two electronic states are $E_1 = 3.94$ eV and $E_2 = 4.84$ eV. The diagonal linear coupling terms of first two modes { $\hat{R}_1$, $\hat{R}_2$ } are $\kappa_1^{(1)} = 0.037$ eV, $\kappa_2^{(1)} = -0.105$ eV for the first electronic state, and $\kappa_1^{(2)} = -0.254$ eV, $\kappa_2^{(2)} = 0.149$ eV for the second electronic state, respectively. The off-diagonal linear coupling of third mode $\hat{R}_3$ is $\lambda_3^{(12)} = \lambda_3^{(21)} = 0.262$ eV. The normal-mode vibronic frequency of each mode is $\omega_1 = 0.126$ eV, $\omega_2 = 0.074$ eV, and $\omega_3 = 0.118$ eV, respectively. Initial conditions of nuclear DOFs are sampled from the corresponding Wigner function of the vibronic ground state while the second electronic state is occupied. All simulations employ $\sim 10^5$ trajectories and time stepsize $\Delta t = 0.01$ fs for fully converged results. Numerically exact result of this model calculated by ML-MCTDH are available in Reference [214].

Figure 9 shows population dynamics of state 2 yielded by wMM, CMM, Ehrenfest dynamics, FSSH and ML-MCTDH. It is evident that Ehrenfest dynamics performs poorly even for the short-time behaviour (before 100 fs). In comparison, wMM, CMM, and FSSH more reasonably describe the radiationless energy transfer process at short time. Interestingly, wMM describes the oscillating behaviours in the long-time region (after 300 fs) better than other approximate methods. Such oscillating behaviour in population dynamics indicates the molecular system passes through the "slopped" conical intersection region[303].

Figures 6-9 demonstrate that the overall performance of wMM is better than CMM, especially in the gas phase scattering case of Figure 7c and the quantum electrodynamic light-matter systems of Figure 8. Both wMM and CMM approaches are able to outperform Ehrenfest dynamics as well as FSSH for condensed phase systems (e.g., in Figure 6 and Figure 8).

## 5. CONCLUSION REMARKS

The phase space formulation of quantum mechanics not only presents a type of convenient interpretation to describe quantum-classical correspondences as well as nonclassical correlations/entanglement, but also sets the insightful scene for developing practical and useful trajectory-based quantum dynamics approaches.

In the Focus Article we show that the constraint coordinate-momentum phase space formulation for the discrete-variable system which we have recently developed, and the weighted representation that we propose in the Focus Article are useful approaches for illustration of nonclassical features of quantum systems. The novel formulation is expected to have potential use for illustration of nonclassical features of quantum states, as well as for future phase point measurement experiment[70, 85, 223-236].

It is straightforward to show the relation between the SU(F)/U(F-1) Stratonovich phase space[114] and constraint coordinate-momentum phase space, which is diffeomorphic to U(F)/U(F-1). When $F > 2$, it is inevitable to meet singularities in dynamics for discrete-variable systems when Stratonovich phase space is used (based on the symplectic structure of the phase space[304]). In comparison, (weighted) constraint coordinate-momentum phase space does not cause any singularities in trajectory-based exact dynamics, which is much more numerically favourable. (See more discussion in Appendix 3 of the Supporting Information).

When the general Moyal bracket of the quantum Liouville theorem is approximated by the corresponding Poisson bracket[57, 58] on (weighted) constraint phase space, it reproduces the correct frozen-nuclei limit of composite/nonadiabatic systems. Such trajectory-based EOMs on (weighted) constraint coordinate-momentum phase space do not rely on the choice of representation of electronic states and are straightforward to obtain the form under covariant transformations. Because second-order nonadiabatic coupling terms are avoided in the EOMs of the adiabatic representation, it is especially useful for applications to realistic molecular systems. (In addition to Section 4.1 , more discussion on the EOMs is presented in Appendices 2 and 5 of the Supporting Information.) Various benchmark model tests of from gas phase to condensed phase quantum systems. (as shown in Figures 6-9) indicate that wMM, the new trajectory-based approximate approach with the weighted constraint coordinate-momentum phase space representation, demonstrates overall better performance than FSSH as well as Ehrenfest dynamics. It is expected that more investigations on the (weighted) constraint phase space formulation will shed light on more numerically favourable dynamics approaches with the Meyer-Miller mapping Hamiltonian or other mapping Hamiltonians (e.g., those of Reference [133] and discussed in Reference [58]).

We note that the (weighted) constraint coordinate-momentum phase space formulation is established for any systems with a finite set of states, not only limited to discrete electronic states, but also for finite discrete nuclear states. The weighted phase space strategy that we propose can also be applied to other types of phase space formulations of the discrete-variable system, such as Stratonovich phase space, and Wootters phase space, albeit that the general coordinate-momentum phase space formulation presented in the Focus Article will be more convenient, for experimental measurements, tomography, or characterizations of fidelity, coherence, inequalities, displaced parity, atomic/molecular/optical Schrodinger cat states, and entanglement in quantum information and computation[70, 85, 223-236, 305, 306] as well as for studying dynamic processes of composite systems in physics, chemistry, materials, biology, and environmental science.

**Figures and Tables**

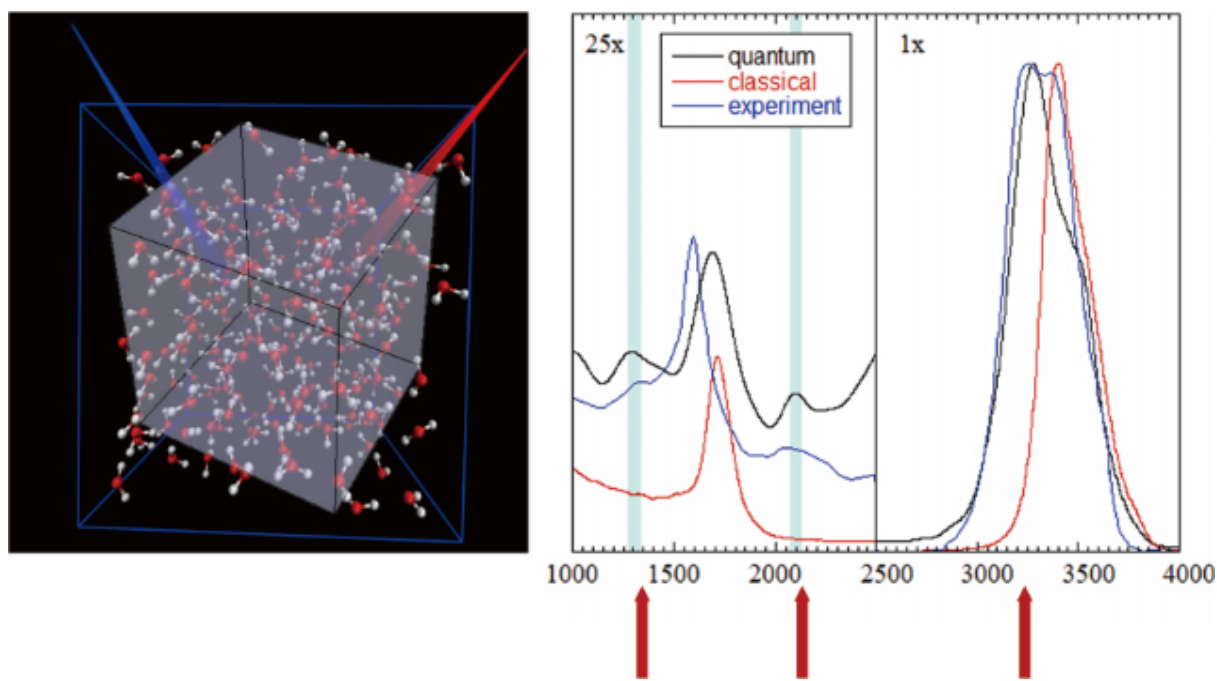

**Figure 1.** Quantum dynamical effects are decisive in reproducing the experimental isotropic Raman spectrum of liquid water at room temperature, as illustrated by the LSC-IVR simulation where infinite (Wigner) phase space for nuclear DOFs is used. Converged results were obtained with 216 water molecules in a box with periodic boundary conditions. (Reprinted with permission from Reference [19]. Copyright 2018 Taylor & Francis.)

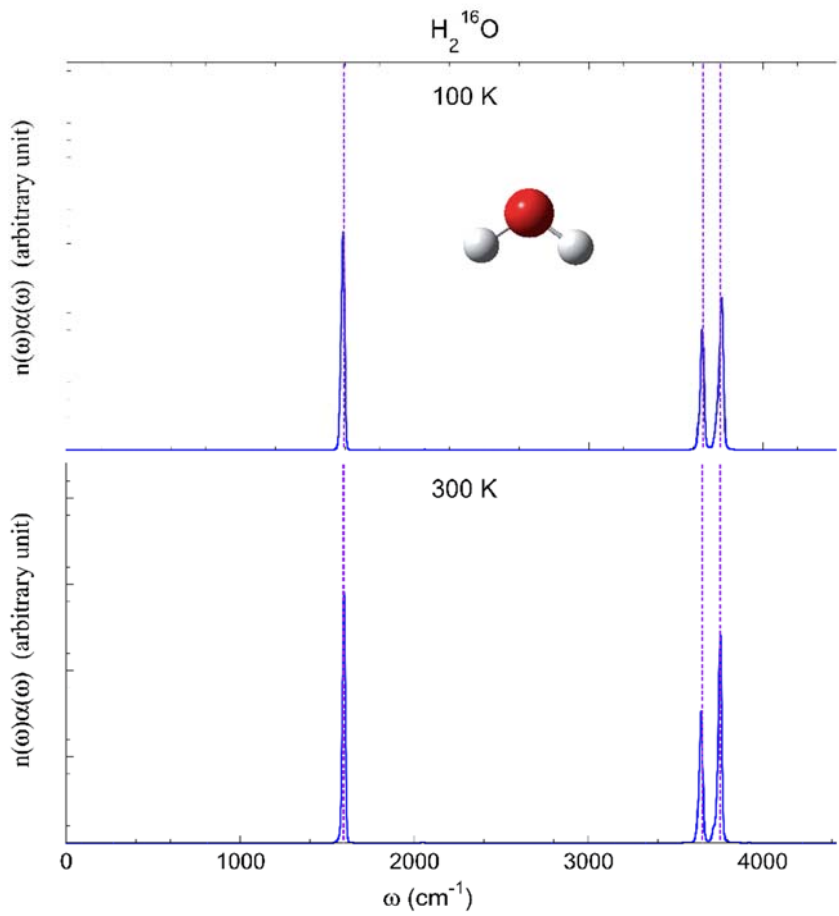

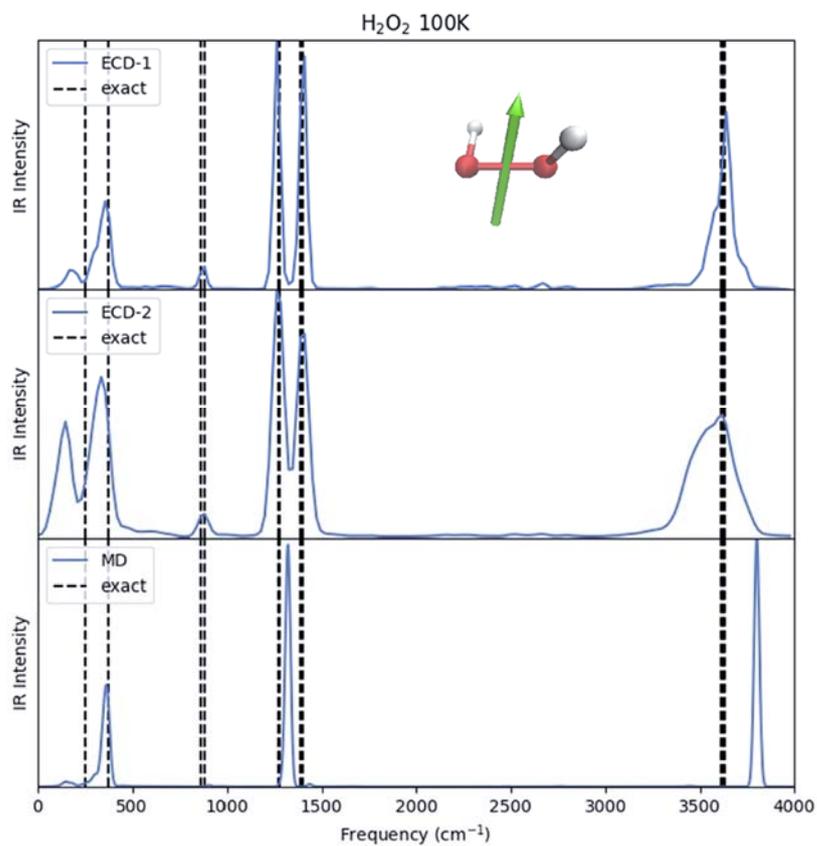

**Figure 2**.  Molecular vibrational spectra produced by more advanced trajectory-based dynamics methods with infinite (Wigner) phase space used for nuclear DOFs, which satisfy the two fundamental criteria: conservation of the quantum Boltzmann distribution for the thermal equilibrium system and being exact for any quantum thermal correlation functions in the classical and harmonic limits.  (a) Vibrational spectrum of the $H_2O$ molecule at 100K and that at 300K.  Adapted with permission from Reference [41].  Copyright 2016 American Institute of Physics Publishing. (b) Vibrational spectrum of the $H_2O_2$ molecule at 100K.  Adapted with permission from Reference [44].  Copyright 2021 American Institute of Physics Publishing.

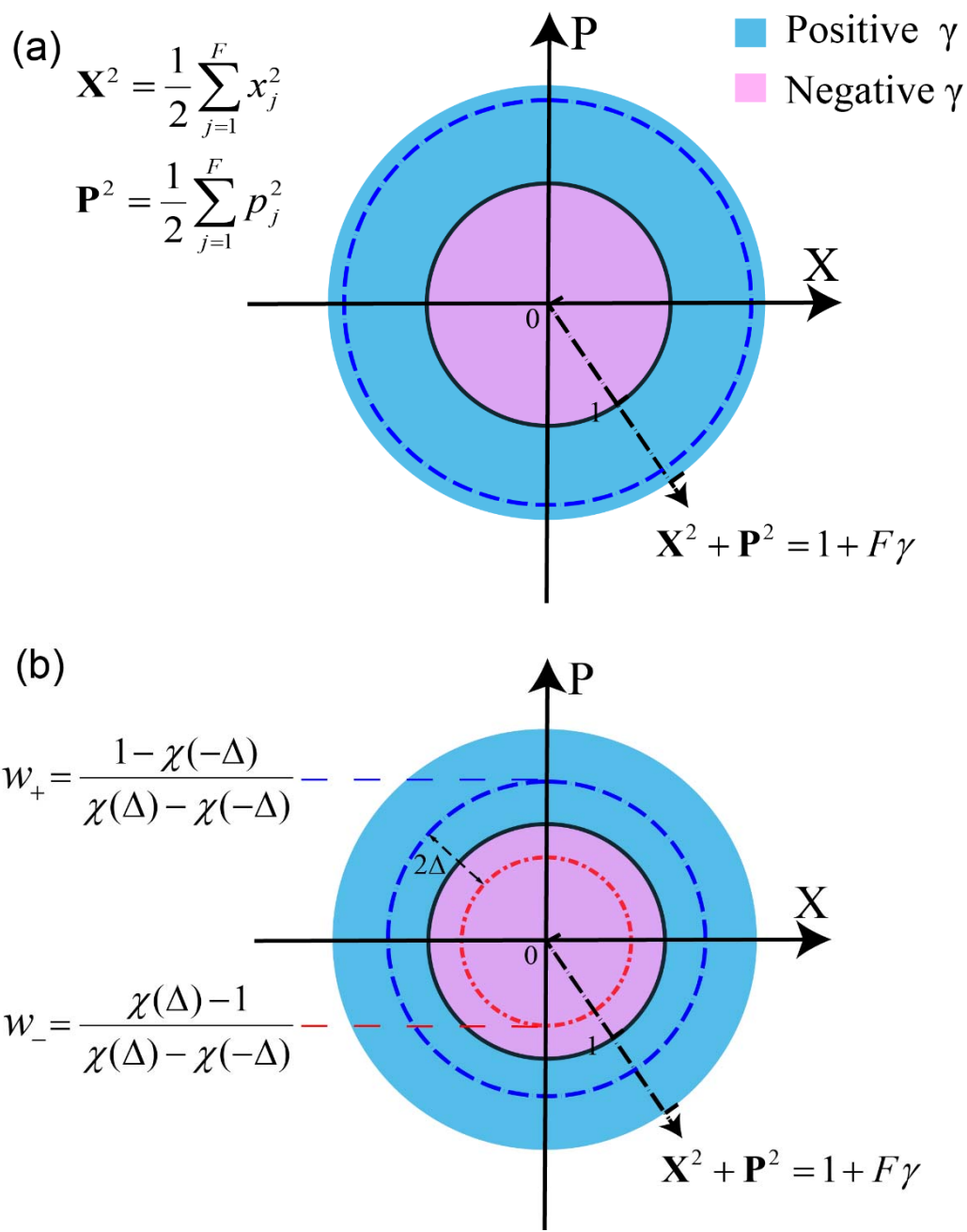

**Figure 3:** Illustration of the exact mapping formulation with constraint coordinate-momentum phase space. Panel (a) presents constraint phase space with only a single value of parameter $\gamma$. Panel (b) demonstrates weighted constraint phase space with two values of parameter $\gamma$, where the quasi-probability distribution function is $w(\gamma) = w_+ \delta(\gamma - \Delta) + w_- \delta(\gamma + \Delta)$. Constraint phase space with the positive weight is blue-dashed, while that with the negative weight is red dot-dashed. (Panel (a) is adapted with permission from Reference [136]. Copyright 2021 American Chemical Society.)

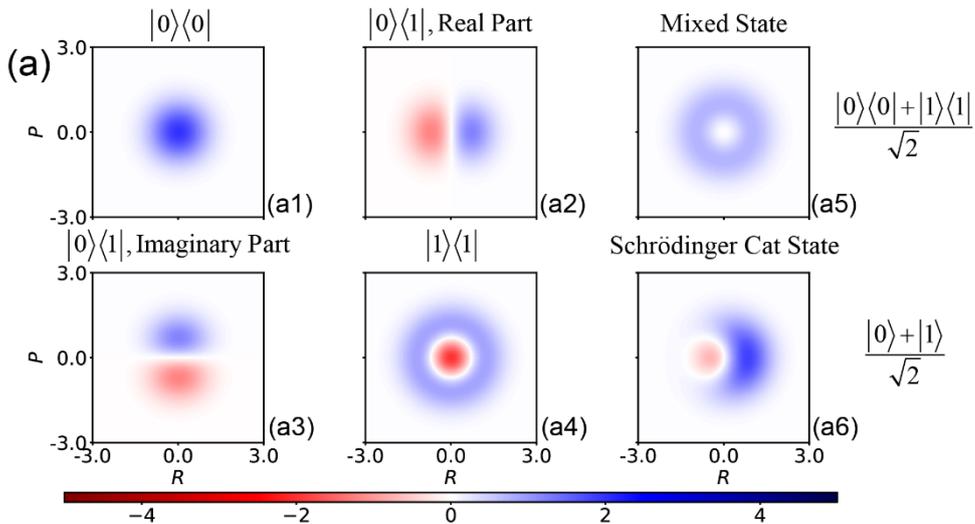
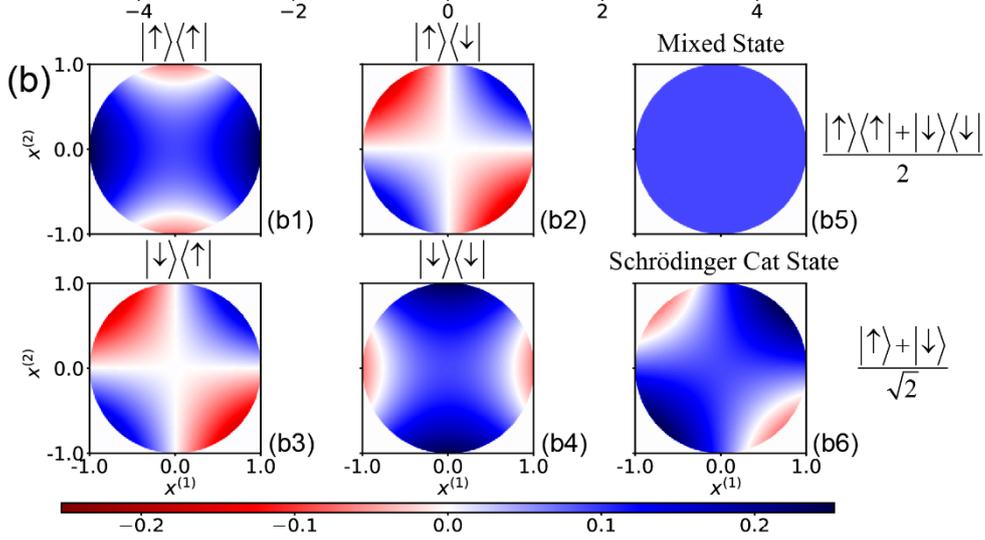
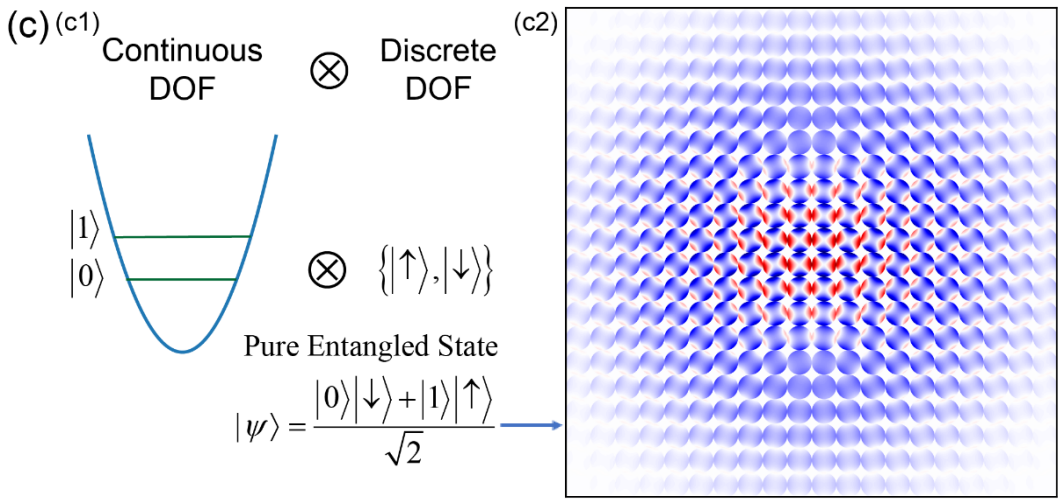

**Figure 4:** Illustrations of (a) Wigner representation of a continuous-variable system, (b) constraint phase space representation of a discrete-variable system, and (c) hybrid coordinate-momentum phase space representation of a composite system with both discrete and continuous DOFs.

(a) Wigner distribution for $|0\rangle\langle 0|$ (Panel a1), that for $|1\rangle\langle 1|$ (Panel a4), real part (Panel a2) and imaginary part (Panel a3) of the Wigner distribution for $|0\rangle\langle 1|$, Wigner distribution for mixed state $(|0\rangle\langle 0|+|1\rangle\langle 1|)/2$ (Panel a5), and that for Schrödinger cat state $(|0\rangle+|1\rangle)/\sqrt{2}$ (Panel a6). Here, $|0\rangle$ and $|1\rangle$ are two energy levels of a continuous-variable system.

(b) Marginal distribution of constraint phase space coordinates $(x^{(1)}, x^{(2)})$ for $|\uparrow\rangle\langle\uparrow|$ (Panel b1), $|\downarrow\rangle\langle\downarrow|$ (Panel b4), that for $|\uparrow\rangle\langle\downarrow|$ (Panel b2), that for $|\downarrow\rangle\langle\uparrow|$ (Panel b3), that for mixed state $(|\uparrow\rangle\langle\uparrow|+|\downarrow\rangle\langle\downarrow|)/2$ (Panel b5), and that for Schrödinger cat state $(|\uparrow\rangle+|\downarrow\rangle)/\sqrt{2}$ (Panel b6). Here, $|\uparrow\rangle$ and $|\downarrow\rangle$ represent two discrete states of a discrete-variable system.

(c) Panel c1: Schematic representaton of the composite system and the pure entangled state $(|0\rangle|\downarrow\rangle+|1\rangle|\uparrow\rangle)/2$; Panel c2: hybrid coordinate-momentum phase space representation of the entangled state. The grid is on the Wigner phase space $(R, P)$ for the continous DOF, and each circle of a grid stands for the local marginal distribution function of constraint phase space variables $(x^{(1)}, x^{(2)})$. The notations are identical to those in Panels (a)-(b).

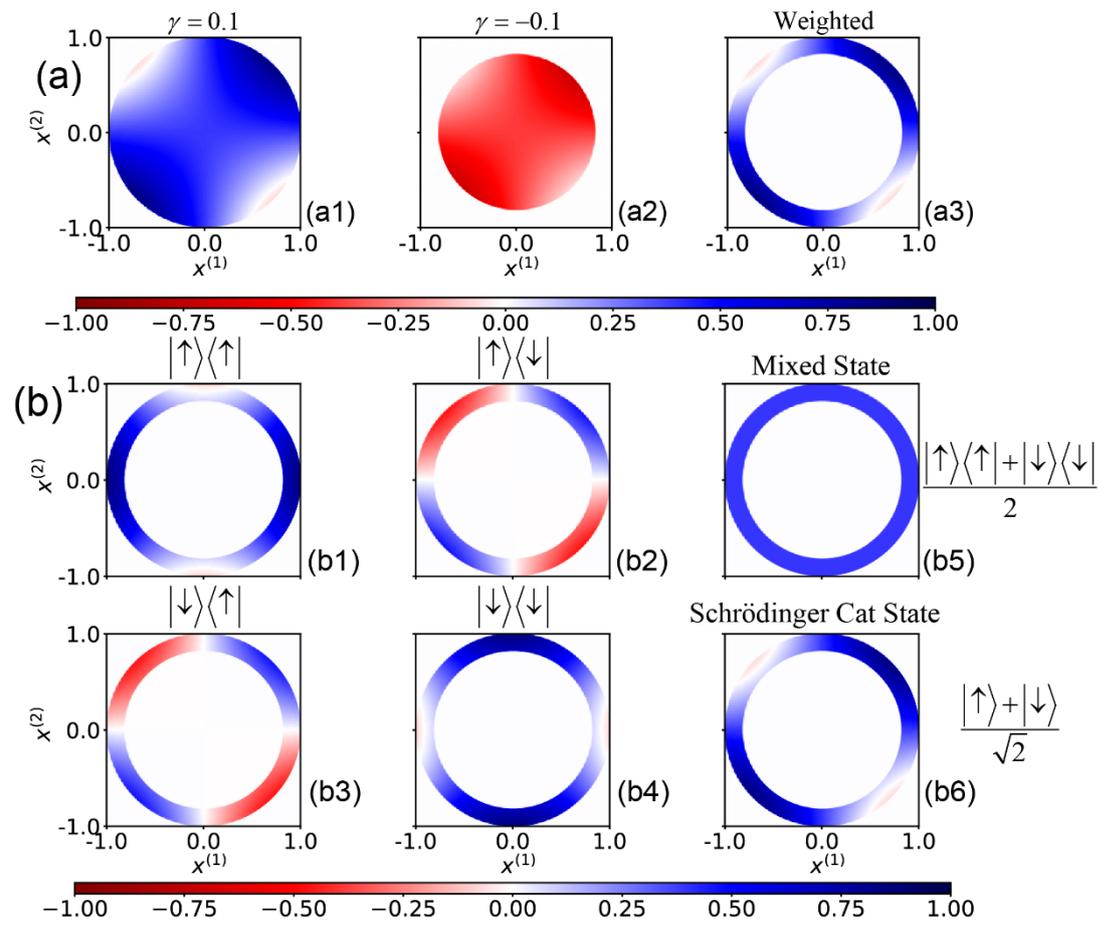
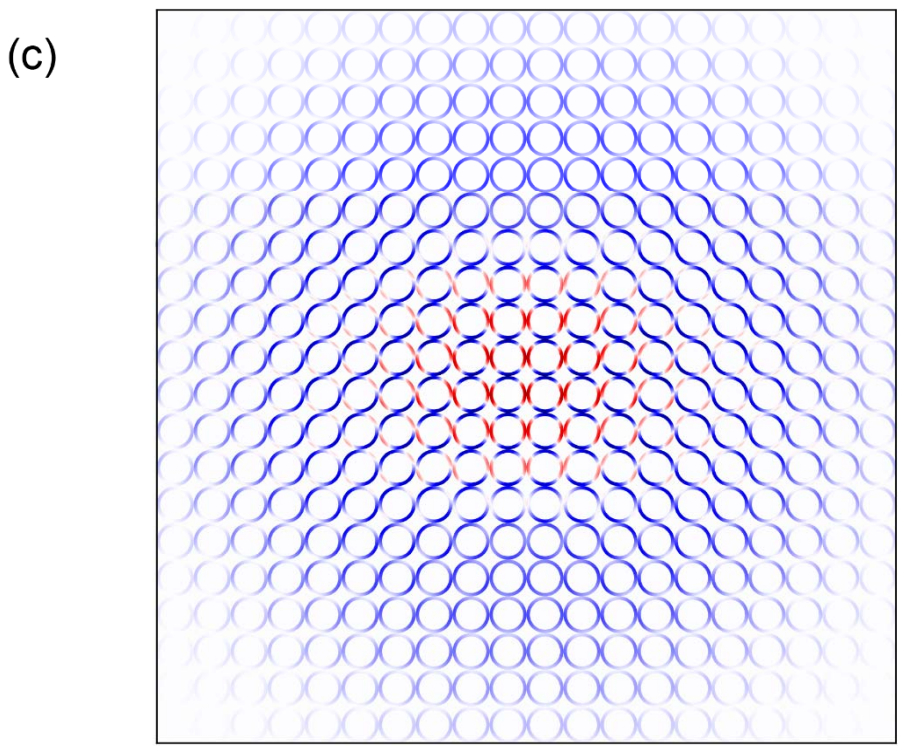

**Figure 5:** Illustrations of (a) components and (b) marginal distrubution functions of the weighted constraint phase space representation of a discrete-variable system, and (c) weighted hybrid representation of the same composite system as that of Figure 4(c).

(a) Marginal distribution of constraint phase space coordinates $(x^{(1)}, x^{(2)})$ for Schrödinger cat state $(|\uparrow\rangle + |\downarrow\rangle)/\sqrt{2}$ with $\gamma = \Delta$ weighted by $w_+$ (Panel a1), with $\gamma = -\Delta$ weighted by $w_-$ (Panel a2). The sum of the two components yields the marginal distribution of constraint phase space coordinates $(x_1, x_2)$ of the weighted representation with two values of parameter $\gamma$ for the Schrödinger cat state (Panel a3). Coordinates are scaled by the larger radius $\sqrt{2(1+2\Delta)}$.

(b) Weighted marginal distribution of constraint phase space coordinates $(x^{(1)}, x^{(2)})$ for the same properties as those in Figure 4(b).

(c) Same as Figure 4(c), but using weighted marginal distribution for the discrete DOF.

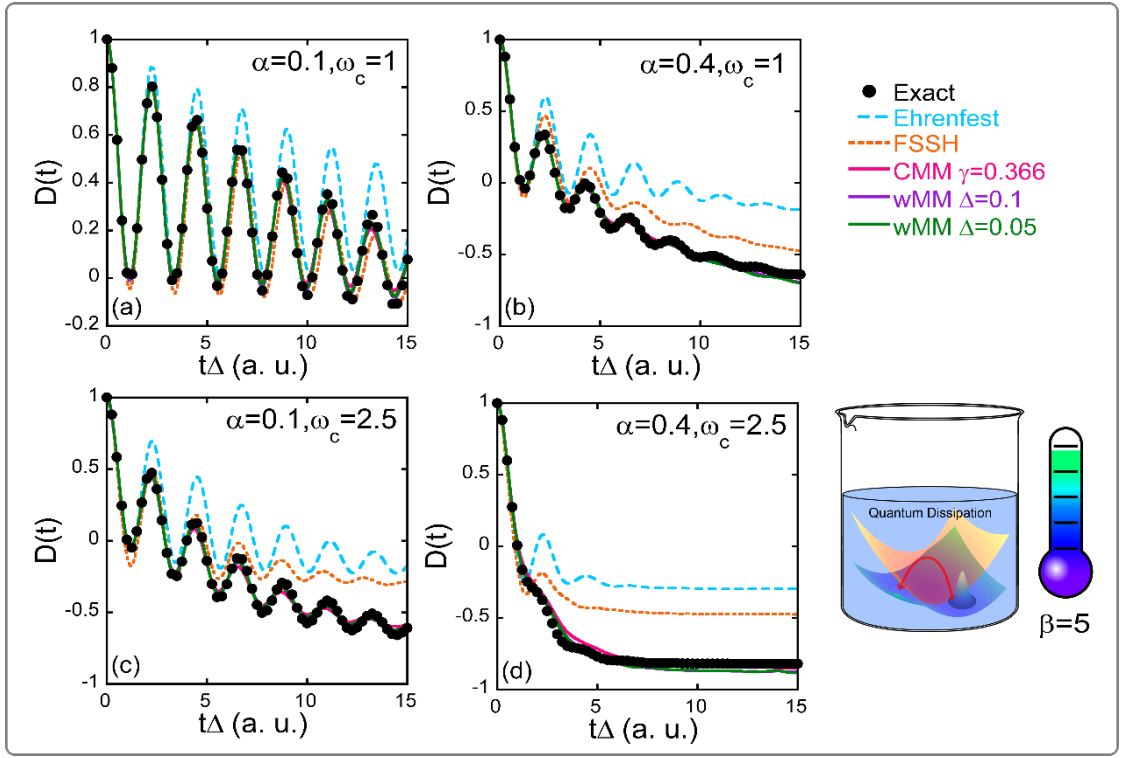

**Figure 6:** Results of population difference $D(t) = P_1(t) - P_0(t)$ between two states for the spin-boson model at low-temperature ($\beta = 1/(k_B T) = 5$) with the Ohmic bath. Panel (a) reports the population dynamics of the spin-boson model with parameters $\varepsilon = \Delta_c = 1$, $\beta = 5$, $\omega_c = 1$, $\alpha = 0.1$ in Panel (a). Solid circles: Exact results produced by eHEOM reported in Reference [136]. Cyan dashed lines: Ehrenfest dynamics. Orange dashed lines: FSSH. Magenta solid lines: CMM with $\gamma = 0.366$. Purple and green solid lines: wMM with $\Delta = 0.1$ and 0.05, respectively. Panel (b) is similar to Panel (a) but for $\alpha = 0.4$; Panel (c) is similar to Panel (a) but for $\omega_c = 2.5$; Panel (d) is similar to Panel (a) but for $\omega_c = 2.5$, $\alpha = 0.4$. In each model 300 continuous DOFs (i.e., effective bath modes) are used.

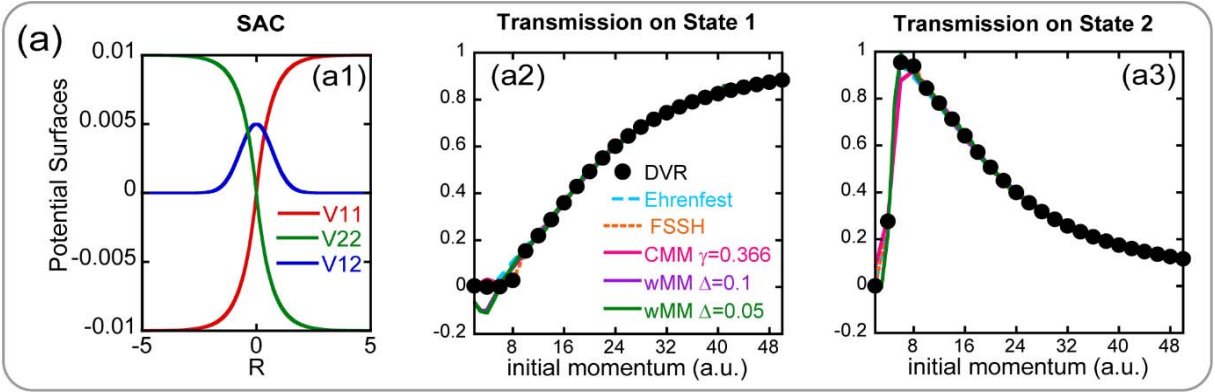
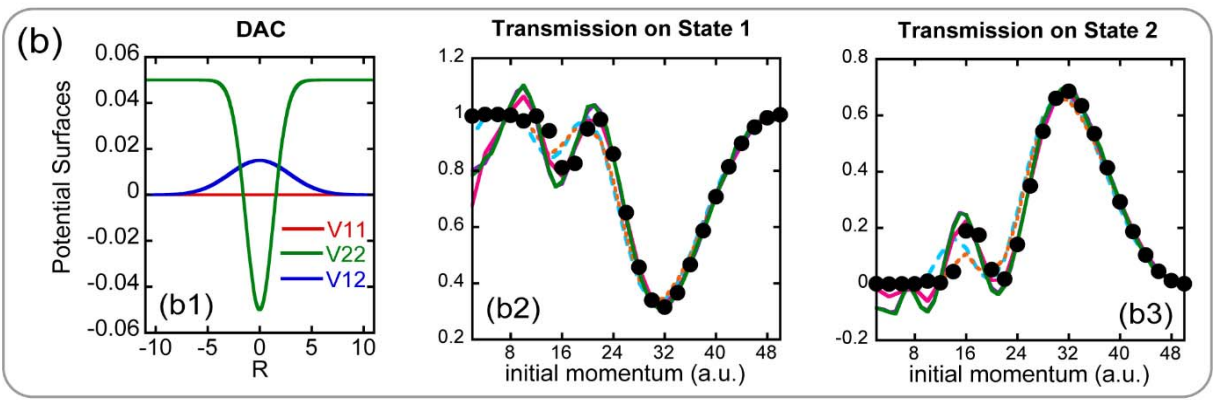
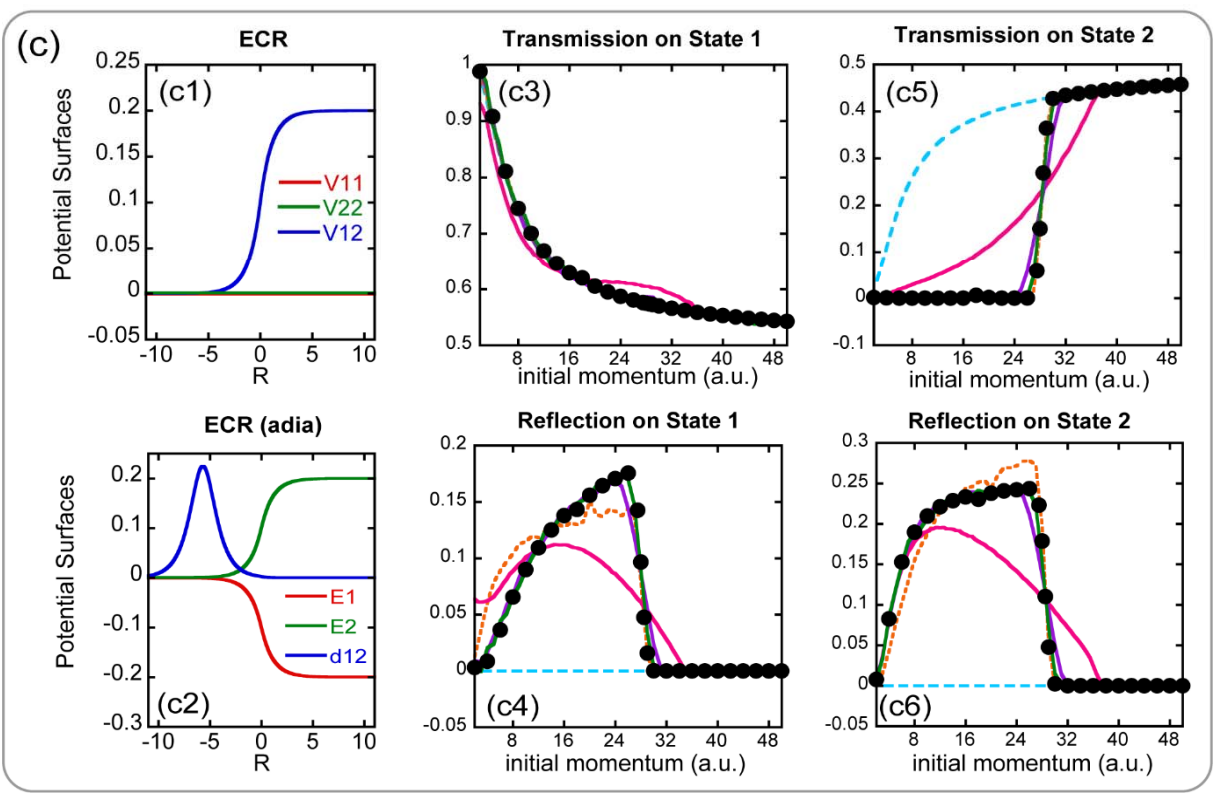

**Figure 7:** Illustration of three Tully models and simulation results. Panel (a1) denotes diabatic PESes $V_{11}(R)$ and $V_{22}(R)$, as well as coupling term $V_{12}(R)$ for the SAC model; Panel (b1) does so for the DAC model; Panel (c1) does so for the ECR model. Panel (c2) demonstrates adiabatic PESes $E_1(R)$ and $E_2(R)$, as well as nonadiabatic coupling vector $d_{12}(R)$.

Panels (a2)-(a3): transmission coefficients on diabatic state 1, and those on diabatic state 2 of the SAC model, respectively. Panels (b2)-(b3): similar to Panels (a2)-(a3), but for the DAC model. Panels (c3) and (c4): transmission/reflection coefficients on adiabatic state 1 of the ECR model; Panels (c5) and (c6): those on adiabatic state 2.

In Panels (a2)-(a3), (b2)-(b3), and (c3)-(c6), magenta, purple and green lines stand for transmission coefficients results for CMM with $\gamma = 0.366$, wMM with $\Delta = 0.1$, and wMM with $\Delta = 0.05$, respectively. Long-dashed blue lines: Ehrenfest dynamics; Short-dashed orange lines: FSSH; Black points: exact DVR benchmarks.

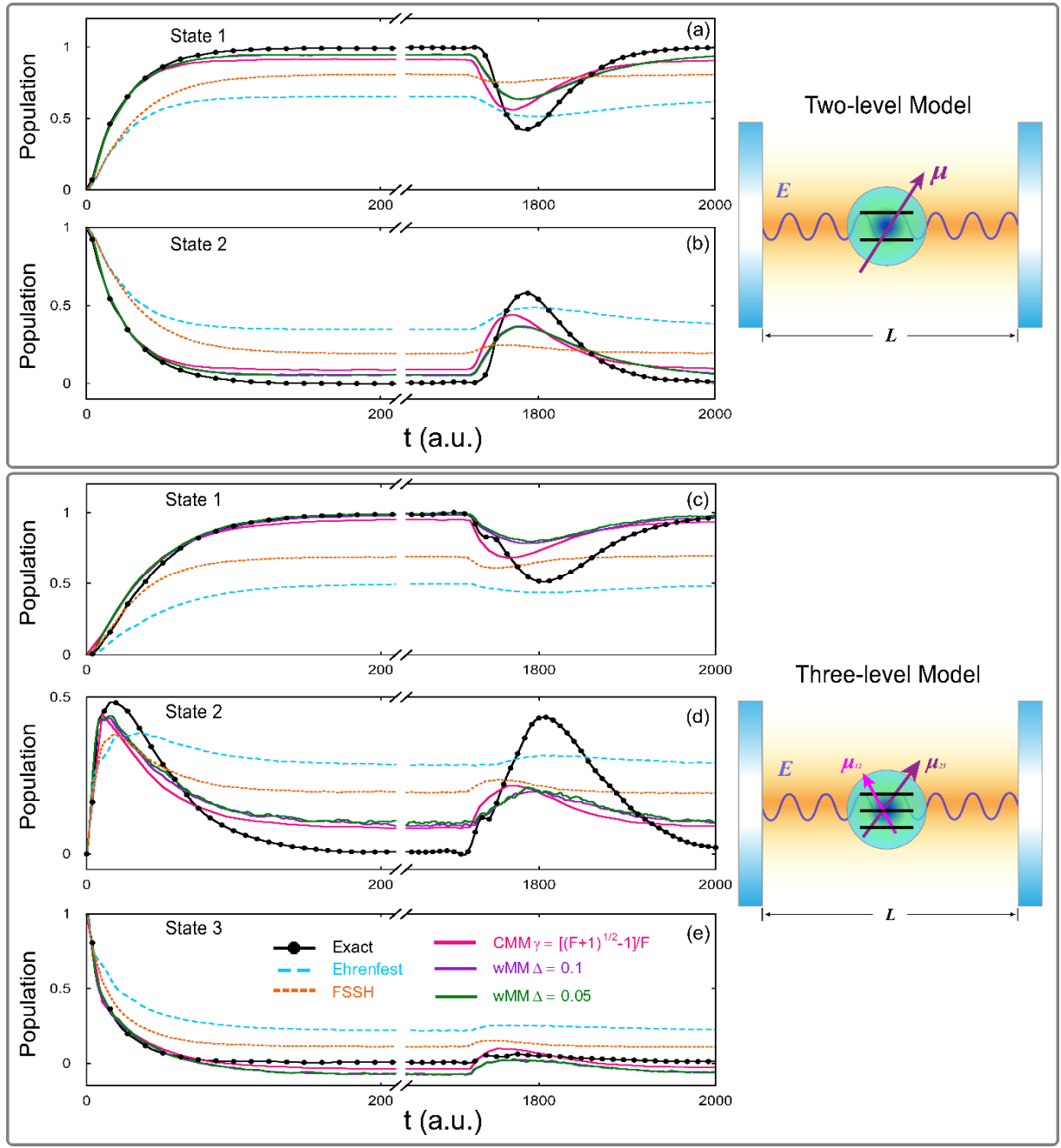

**Figure 8:** Results of population dynamics for the atom-in-cavity models. Panels (a)-(b) represent data of the first and second states of the two-level model, respectively. Panels (c)-(e) denote data of the first, second and third states of the three-level model, respectively. Magenta solid lines: CMM with $\gamma = \left(\sqrt{F+1}-1\right)/F$; Purple solid lines: wMM with $\Delta = 0.1$; Green solid lines: wMM with $\Delta = 0.05$; Cyan long-dashed lines: Ehrenfest dynamics; Orange short-dashed lines: FSSH; Black solid-dotted lines: exact results from References [296, 297]. In each model 400 continuous DOFs (i.e., standing-wave modes) are involved.

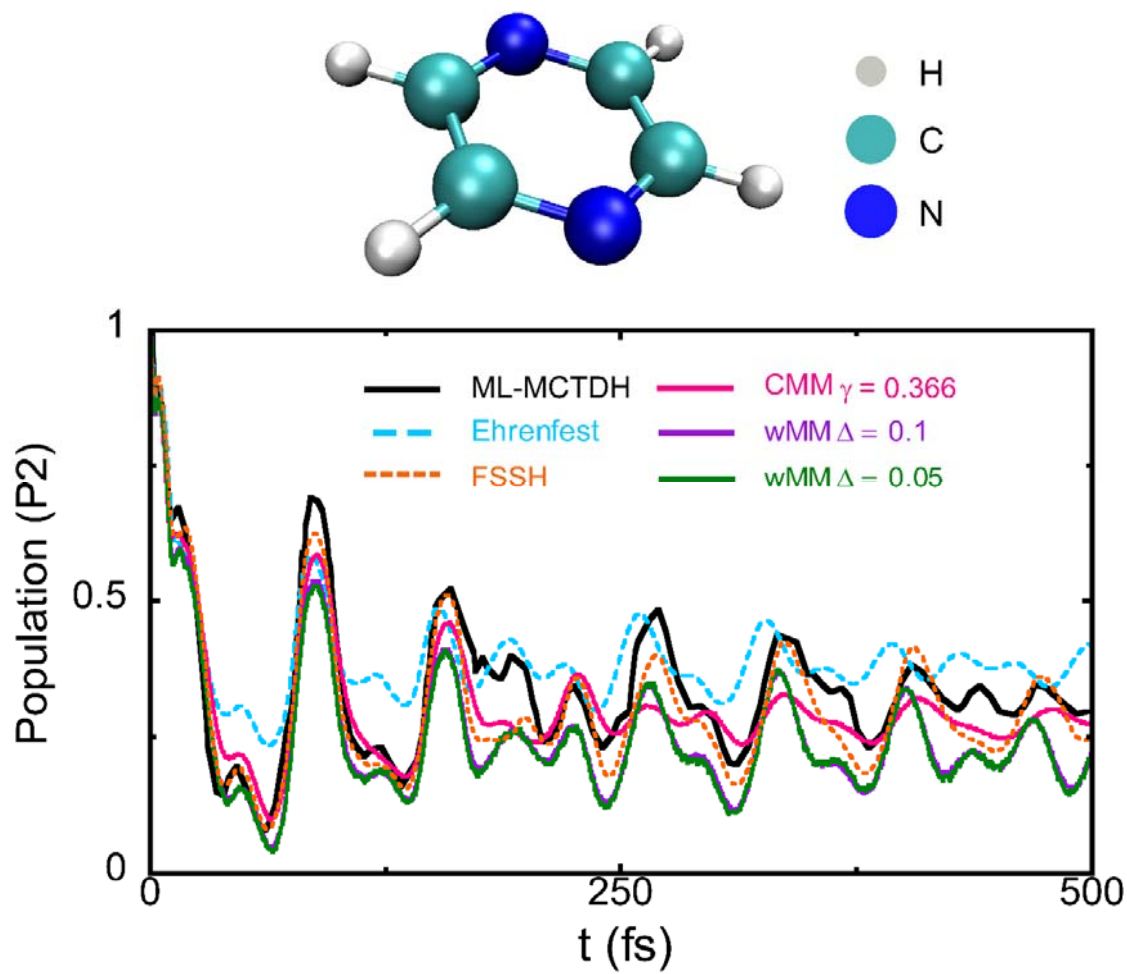

**Figure 9:** Results of population dynamics of the second electronic state of the 2-level 3-mode pyrazine model. Magenta solid lines: CMM with $\gamma = \left(\sqrt{F+1}-1\right)/F \approx 0.366$; Purple solid lines: wMM with $\Delta = 0.1$; Green solid lines: wMM with $\Delta = 0.05$. Cyan dashed lines: Ehrenfest dynamics; Orange short-dashed lines: FSSH; Black solid lines: ML-MCTDH results of Reference [214].

**Funding Information**

National Natural Science Foundation of China (NSFC) Grant No. 21961142017; Ministry of Science and Technology of China (MOST) Grant No. 2017YFA0204901

**Research Resources**

High-performance Computing Platform of Peking University, Beijing PARATERA Tech CO., Ltd., and Guangzhou supercomputer center

**Acknowledgments**

This work was supported by the National Natural Science Foundation of China (NSFC) Grant No. 21961142017, and by the Ministry of Science and Technology of China (MOST) Grant No. 2017YFA0204901. We acknowledge the High-performance Computing Platform of Peking University, Beijing PARATERA Tech CO., Ltd., and Guangzhou supercomputer center for providing computational resources.

**Notes**

The authors declare no competing financial interest.

**Supporting Information**

**Appendix 1. Derivation of the formulation of (weighted) constraint coordinate-momentum phase space.**

In this appendix, we introduce the techniques for performing integrals on constraint coordinate-momentum phase space, and show the brief derivation of the formulation of (weighted) constraint coordinate-momentum phase space. Such techniques have been expressed in Appendix A of ref [134] as well as in Section S1 of the Supporting Information of ref [136].

Consider the expectation of any function $f(X_1, X_2, \cdots, X_L)$ under the *L*-dimensional independent standard normal distribution (i.e., $\{X_i\} \sim N^{(L)}(0,1)$). It is closely related to that under the uniform distribution on the (*L*-1)-dimensional constraint space, i.e.,

$$\langle f \rangle_{N^{(L)}(0,1)} = \frac{1}{(2\pi)^{L/2}} \int dX_1 dX_2 \cdots dX_L e^{-\frac{1}{2}\sum_{k=1}^{L} X_k^2} f(X_1, X_2, \cdots, X_L)$$

$$= \frac{1}{(2\pi)^{L/2}} \int_0^\infty d\xi \int dX_1 dX_2 \cdots dX_L e^{-\xi} \delta\left(\frac{1}{2}\sum_{k=1}^{L} X_k^2 - \xi\right) f(X_1, X_2, \cdots, X_L) \quad . \quad (86)$$

$$= \frac{1}{(2\pi)^{L/2}} \int_0^\infty d\xi e^{-\xi} S_{L-1}\left(\sqrt{\xi}\right) \langle f \rangle_{S_{L-1}(\sqrt{\xi})}$$

Here,

$$S_{L-1}(R) = \int dX_1 dX_2 \cdots dX_L \delta\left(\frac{1}{2}\sum_{k=1}^{L} X_k^2 - R^2\right) \tag{87}$$

denotes the area of the constraint space with radius $R$, and the expectation under the uniform distribution on the (L-1)-dimensional constraint space, $\langle \cdots \rangle_{S_{L-1}(R)}$, is given by

$$\langle f \rangle_{S_{L-1}(R)} = \frac{1}{S_{L-1}(R)} \int dX_1 dX_2 \cdots dX_L \delta\left(\frac{1}{2}\sum_{k=1}^{L} X_k^2 - R^2\right) f(X_1, X_2, \cdots, X_L) \quad . \tag{88}$$

(Note that the constraint space with radius $R$ defined in eq (87) is slightly different from the (L-1)-sphere in mathematics.) It is trivial to verify that $S_{L-1}(R) = R^{L-2} S_{L-1}(1)$ is the scaling relation of $S_{L-1}(R)$. We can calculate $S_{L-1}(1)$ by setting $f(X_1, X_2, \cdots, X_L) \equiv 1$ in eq (86), i.e.,

$$1 = \frac{1}{(2\pi)^{L/2}} \int_0^\infty d\xi e^{-\xi} S_{L-1}\left(\sqrt{\xi}\right) = \frac{1}{(2\pi)^{L/2}} S_{L-1}(1) \int_0^\infty d\xi e^{-\xi} \xi^{\frac{L-2}{2}} \quad . \tag{89}$$

Equation (89) yields

$$S_{L-1}(1) = \frac{(2\pi)^{L/2}}{\Gamma(L/2)} \tag{90}$$

after we use the equality $\int_0^\infty de^{-x} x^\alpha = \Gamma(\alpha+1)$ for $\alpha > -1$, where the Gamma function is defined by $\Gamma(s) = \int_0^\infty dt e^{-t} t^{s-1} \ (s>0)$.

The *k*-th order moment of the uniform distribution on the (*L*-1)-dimensional constraint space has the scaling relation,

$$\left\langle X_{n_1} X_{n_2} \cdots X_{n_k} \right\rangle_{S_{L-1}(\sqrt{\xi})} = \frac{1}{S_{L-1}(\sqrt{\xi})} \int dX_1 dX_2 \cdots dX_L \delta\left(\frac{1}{2}\sum_{l=1}^{L} X_l^2 - \xi\right) X_{n_1} X_{n_2} \cdots X_{n_k}$$
$$= \xi^{k/2} \left\langle X_{n_1} X_{n_2} \cdots X_{n_k} \right\rangle_{S_{L-1}(1)} \quad (91)$$

Substitution of eq (91) into eq (86) produces

$$\left\langle X_{n_1} X_{n_2} \cdots X_{n_k} \right\rangle_{N^{(L)}(0,1)} = \frac{\Gamma\left(\frac{L+k}{2}\right)}{\Gamma(L/2)} \left\langle X_{n_1} X_{n_2} \cdots X_{n_k} \right\rangle_{S_{L-1}(1)}$$
$$= \frac{\Gamma\left(\frac{L+k}{2}\right)}{\xi^{k/2} \Gamma(L/2)} \left\langle X_{n_1} X_{n_2} \cdots X_{n_k} \right\rangle_{S_{L-1}(\sqrt{\xi})} \quad (92)$$

The well-known Wick theorem (also called Isserlis's theorem)[244, 245] for the independent and identical standard normal distribution reads

$$\left\langle X_{n_1} X_{n_2} \cdots X_{n_k} \right\rangle_{N^{(L)}(0,1)} = \begin{cases} 0, & k \text{ is odd} \\ \sum\left(\prod \left\langle X_{n_i} X_{n_j} \right\rangle_{N^{(L)}(0,1)}\right), & k \text{ is even} \end{cases} \quad (93)$$

Here, the notation $\sum\left(\prod \cdots\right)$ stands for the summation of the products over all possible pair partitions of $\{X_{n_1}, X_{n_2}, \cdots, X_{n_k}\}$. It is straightforward to obtain the second order moment and fourth order moment of $N^{(L)}(0,1)$,

$$\begin{aligned} \left\langle X_i X_j \right\rangle_{N(0,1)} &= \delta_{ij} \\ \left\langle X_i X_j X_k X_l \right\rangle_{N(0,1)} &= \delta_{ij}\delta_{kl} + \delta_{ik}\delta_{jl} + \delta_{il}\delta_{jk} \end{aligned} \quad (94)$$

Substitution of eq (94) into eq (92) leads to

$$\langle X_i X_j \rangle_{S_{L-1}(\sqrt{\xi})} = \frac{\xi}{L/2}\delta_{ij}$$

$$\langle X_i X_j X_k X_l \rangle_{S_{L-1}(\sqrt{\xi})} = \frac{\xi^2}{\frac{L}{2}\left(\frac{L}{2}+1\right)}\left(\delta_{ij}\delta_{kl} + \delta_{ik}\delta_{jl} + \delta_{il}\delta_{jk}\right) \quad . \tag{95}$$

We then set $L = 2F$ as well as $\xi = 1 + F\gamma$ and relabel the constraint space (defined by eq 31) as $\mathcal{S}_\gamma(\mathbf{x}, \mathbf{p})$ parameterized by $\gamma$. The comparison between eq 32 and eq (87) states that the area of constraint phase space is

$$\Omega(\gamma) = S_{2F-1}\left(R = \sqrt{1+F\gamma}\right) = \frac{(2\pi)^F}{\Gamma(F)}(1+F\gamma)^{F-1} \quad . \tag{96}$$

We then obtain

$$\langle g(\mathbf{x}, \mathbf{p}) \rangle_{S_{L-1}(\sqrt{\xi})} = \int_{\mathcal{S}_\gamma(\mathbf{x},\mathbf{p})} d\mathbf{X}\, g(\mathbf{x}, \mathbf{p})$$
$$= \int F d\mathbf{x} d\mathbf{p} \frac{1}{\Omega(\gamma)} \delta\left(\sum_{n=1}^{F}\frac{(x^{(n)})^2 + (p^{(n)})^2}{2} - (1+F\gamma)\right) g(\mathbf{x}, \mathbf{p}) \quad , \tag{97}$$

where $\mathbf{X} = (\mathbf{x}, \mathbf{p})$. The second order moment and fourth order moment on the constraint space of eq 31 read

$$\int_{\mathcal{S}_\gamma(\mathbf{x},\mathbf{p})} d\mathbf{X}\left(X_i X_j\right) = \frac{1+F\gamma}{F}\delta_{ij}$$

$$\int_{\mathcal{S}_\gamma(\mathbf{x},\mathbf{p})} d\mathbf{X}\left(X_i X_j X_k X_l\right) = \frac{(1+F\gamma)^2}{F(F+1)}\left(\delta_{ij}\delta_{kl} + \delta_{ik}\delta_{jl} + \delta_{il}\delta_{jk}\right) \quad . \tag{98}$$

Consider $\hat{A} = |m\rangle\langle n|$ and $\hat{B} = |k\rangle\langle l|$. We have

$$\text{Tr}\left[\hat{A}\hat{B}\right] = \delta_{nk}\delta_{ml} \quad . \tag{99}$$

The constraint phase space expression (i.e., eqs 3-5 when only discrete electronic DOFs exist) of the left-hand side (LHS) of eq (99) is equivalent to

$$\mathrm{Tr}\left[\hat{A}\hat{B}\right] = F\int_{\mathcal{S}_\gamma(\mathbf{x},\mathbf{p})} d\mathbf{X} \, \langle n|\hat{K}_{\mathrm{ele}}(\mathbf{x},\mathbf{p})|m\rangle\langle l|K_{\mathrm{ele}}^{-1}(\mathbf{x},\mathbf{p})|k\rangle$$
$$= F\int_{\mathcal{S}_\gamma(\mathbf{x},\mathbf{p})} d\mathbf{X} \, K_{nm}^{(\mathrm{ele})}(\mathbf{x},\mathbf{p}) K_{(\mathrm{ele}),lk}^{-1}(\mathbf{x},\mathbf{p}) \quad . \tag{100}$$

In eq (100), $K_{nm}^{(\mathrm{ele})}(\mathbf{x},\mathbf{p})$ is the element of operator $\hat{K}_{\mathrm{ele}}(\mathbf{x},\mathbf{p})$, and $K_{(\mathrm{ele}),lk}^{-1}(\mathbf{x},\mathbf{p})$ is the element of operator $K_{\mathrm{ele}}^{-1}(\mathbf{x},\mathbf{p})$. When the mapping kernel is given by eq 29, it is straightforward to derive that its corresponding inverse kernel is eq 30 by employing eqs (98)-(100). Specifically, when the mapping kernel is identical to its inverse (i.e., eq 38), the element of the (inverse) kernel operator is

$$K_{nm}(\mathbf{x},\mathbf{p}) = K_{nm}^{-1}(\mathbf{x},\mathbf{p}) = \frac{1}{2}\left(x^{(n)} + ip^{(n)}\right)\left(x^{(m)} - ip^{(m)}\right) - \gamma\delta_{nm} \quad . \tag{101}$$

It is trivial to use eq (98) and eq (101) to achieve the equality

$$F\int_{\mathcal{S}_\gamma(\mathbf{x},\mathbf{p})} d\mathbf{X} \, K_{nm}(\mathbf{x},\mathbf{p}) K_{lk}^{-1}(\mathbf{x},\mathbf{p}) = \frac{(1+F\gamma)^2}{(F+1)}\delta_{nk}\delta_{ml} + \left[\frac{(1+F\gamma)^2}{(F+1)} - 2\gamma - F\gamma^2\right]\delta_{lk}\delta_{nm} \quad . \tag{102}$$

After employing eqs (99)-(100), we obtain $\gamma = (\sqrt{F+1}-1)/F$ from eq (102) where the mapping kernel is identical to its inverse (i.e., eq 38).

Similarly, the weighting constraint phase space formulation (eqs 3-5 when only discrete electronic DOFs exist, and eq 37) leads to

$$\mathrm{Tr}_e\left[\hat{A}\hat{B}\right] = \int_{\mathcal{S}(\mathbf{x},\mathbf{p})} F d\mathbf{x}d\mathbf{p} \, A_C(\mathbf{x},\mathbf{p})\tilde{B}_C(\mathbf{x},\mathbf{p})$$
$$= \int_{-1/F}^{\infty} d\gamma \, w(\gamma) \int F d\mathbf{x}d\mathbf{p} \frac{1}{\Omega(\gamma)}\delta\left(\sum_{n=1}^{F}\frac{(x^{(n)})^2+(p^{(n)})^2}{2}-(1+F\gamma)\right)A_C(\mathbf{x},\mathbf{p})\tilde{B}_C(\mathbf{x},\mathbf{p})$$
$$= \int_{-1/F}^{\infty} d\gamma \, w(\gamma) \, \langle A_C(\mathbf{x},\mathbf{p})\tilde{B}_C(\mathbf{x},\mathbf{p})\rangle_{S_{L-1}(\sqrt{\xi})}$$
$$\tag{103}$$

Employing eq (99), we obtain an equation similar to eq (100)

$$F\int d\gamma w(\gamma) \int_{\mathcal{S}_\gamma(\mathbf{x},\mathbf{p})} d\mathbf{X} \, K_{nm}(\mathbf{x},\mathbf{p}) K_{lk}^{-1}(\mathbf{x},\mathbf{p}) = \delta_{nk}\delta_{ml} \quad . \tag{104}$$

Because we request eq 38 that the mapping kernel is identical to its inverse in the weighted constraint phase space formulation, eqs (101)-(102) also hold. Substitution of eq (102) into eq (104) generates

$$\int d\gamma w(\gamma)\left(F\gamma^2 + 2\gamma\right) = \int d\gamma w(\gamma)\chi(\gamma) = 1 \quad , \tag{105}$$

which is eq 39 of the main text. When $w(\gamma)$, the quasi-probability distribution function of parameter $\gamma$, is a single Dirac delta function, the constraint coordinate-momentum phase space formulation with $\gamma = (\sqrt{F+1}-1)/F$ is a special case of eq (105), or of eq 39 of the main text. When in eq (105) or eq 39 $w(\gamma)$ is represented by a linear combination of two symmetrical delta functions (i.e., eqs 43-45), we achieve the symmetrically weighted constraint phase space formulation employed in wMM of the Focus Article. Apparently, various other choices for quasi-probability distribution function $w(\gamma)$ are also available and can be investigated in the future.

**Appendix 2. More discussion on the equations of motion in the adiabatic representation**

Below we will show that eq 74 and eq 76, the equations of motion (EOMs) under the diabatic-to-adiabatic transformation are intrinsically equivalent to Hamilton's EOMs generated by the Hamiltonian, eq 79, when (nuclear) canonical phase variables $(\tilde{\mathbf{R}}, \tilde{\mathbf{P}})$ are defined by eq 78. For simplicity, assume that all nonadiabatic coupling terms $\{\mathbf{d}_{mn}^{(I)}(\mathbf{R})\}$ are real. It leads to a simplified relation

$$\tilde{\mathbf{P}} = \mathbf{P} + \hbar \sum_{n,m=1}^{F} \tilde{x}^{(n)} \tilde{p}^{(m)} \mathbf{d}_{mn}(\mathbf{R}) \tag{106}$$

in eq 78. In the adiabatic representation, $\mathbf{P} = \tilde{\mathbf{P}} - \hbar \sum_{n,m=1}^{F} \tilde{x}^{(n)} \tilde{p}^{(m)} \mathbf{d}_{mn}(\tilde{\mathbf{R}})$ is denoted the kinematic momentum (e.g., in ref [186]), which is different from canonical momentum $\tilde{\mathbf{P}}$. (Yet, $\mathbf{R} = \tilde{\mathbf{R}}$.) The Hamiltonian, eq 79, is recast into

$$H_C(\tilde{\mathbf{R}}, \tilde{\mathbf{P}}, \tilde{\mathbf{x}}, \tilde{\mathbf{p}}) = \sum_{I=1}^{N} \frac{\left(\tilde{P}_I - \hbar \sum_{n,m=1}^{F} \tilde{x}^{(n)} \tilde{p}^{(m)} d_{mn}^{(I)}(\tilde{\mathbf{R}})\right)^2}{2M_I} + \sum_{k=1}^{F} E_k(\tilde{\mathbf{R}})\left(\frac{1}{2}\left(\left(\tilde{x}^{(k)}\right)^2 + \left(\tilde{p}^{(k)}\right)^2\right) - \gamma\right) \quad .\tag{107}$$

Here, $d_{mn}^{(I)}(\tilde{\mathbf{R}})$ and $\tilde{P}_I$ are the the $I$-th DOF component of $\mathbf{d}_{mn}(\tilde{\mathbf{R}})$ and that of $\tilde{\mathbf{P}}$, respectively. Note that canonical variables $(\tilde{\mathbf{R}}, \tilde{\mathbf{P}}, \tilde{\mathbf{x}}, \tilde{\mathbf{p}})$ are independent of one another. Hamilton's EOMs produced by eq (107) are

$$\dot{\tilde{x}}^{(m)} = \frac{1}{\hbar} \frac{\partial}{\partial \tilde{p}^{(m)}} H_C(\tilde{\mathbf{R}}, \tilde{\mathbf{P}}, \tilde{\mathbf{x}}, \tilde{\mathbf{p}})$$
$$= -\sum_{J=1}^{N} \sum_{n=1}^{F} \frac{\tilde{P}_J - \hbar \sum_{k,l=1}^{F} \tilde{x}^{(k)} \tilde{p}^{(l)} d_{lk}^{(J)}(\tilde{\mathbf{R}})}{M_J} \tilde{x}^{(n)} d_{mn}^{(J)}(\tilde{\mathbf{R}}) + \frac{1}{\hbar} E_m(\tilde{\mathbf{R}}) \tilde{p}^{(m)} \quad , \tag{108}$$

$$\dot{\tilde{p}}^{(m)} = -\frac{1}{\hbar} \frac{\partial}{\partial \tilde{x}^{(m)}} H_C(\tilde{\mathbf{R}}, \tilde{\mathbf{P}}, \tilde{\mathbf{x}}, \tilde{\mathbf{p}})$$
$$= \sum_{J=1}^{N} \sum_{n=1}^{F} \frac{\tilde{P}_J - \hbar \sum_{k,l=1}^{F} \tilde{x}^{(k)} \tilde{p}^{(l)} d_{lk}^{(J)}(\tilde{\mathbf{R}})}{M_J} \tilde{p}^{(n)} d_{nm}^{(J)}(\tilde{\mathbf{R}}) - \frac{1}{\hbar} E_m(\tilde{\mathbf{R}}) \tilde{x}^{(m)} \quad , \tag{109}$$

$$\dot{\tilde{R}}_I = \frac{\partial}{\partial \tilde{P}_I} H_C(\tilde{\mathbf{R}}, \tilde{\mathbf{P}}, \tilde{\mathbf{x}}, \tilde{\mathbf{p}}) = \frac{\tilde{P}_I - \hbar \sum_{n,m=1}^{F} \tilde{x}^{(n)} \tilde{p}^{(m)} d_{mn}^{(I)}(\tilde{\mathbf{R}})}{M_I} \quad , \tag{110}$$

$$\dot{\tilde{P}}_I = -\frac{\partial}{\partial \tilde{R}_I} H_C(\tilde{\mathbf{R}}, \tilde{\mathbf{P}}, \tilde{\mathbf{x}}, \tilde{\mathbf{p}})$$
$$= \hbar \sum_{J=1}^{N} \sum_{n,m=1}^{F} \frac{\tilde{P}_J - \hbar \sum_{k,l=1}^{F} \tilde{x}^{(k)} \tilde{p}^{(l)} d_{lk}^{(J)}(\tilde{\mathbf{R}})}{M_J} \tilde{x}^{(n)} \tilde{p}^{(m)} \frac{\partial d_{mn}^{(J)}(\tilde{\mathbf{R}})}{\partial \tilde{R}_I} \quad . \tag{111}$$
$$- \sum_{k=1}^{F} \frac{\partial E_k(\tilde{\mathbf{R}})}{\partial \tilde{R}_I} \left( \frac{1}{2} \left( (\tilde{x}^{(k)})^2 + (\tilde{p}^{(k)})^2 \right) - \gamma \right)$$

It is evident that eqs (108)-(109) are identical to eq 70 and that eq (110) is the same as the first equation of eq 76. We will then prove that eq (111) is equivalent to the second equation of eq 76.

Consider the full time-derivative of canonical momentum $\tilde{P}_I$,

$$\dot{\tilde{P}}_I = \dot{P}_I\left(\tilde{\mathbf{R}},\tilde{\mathbf{P}},\tilde{\mathbf{x}},\tilde{\mathbf{p}}\right) + \hbar \frac{\mathrm{d}}{\mathrm{d}t} \sum_{n,m=1}^{F} \tilde{x}^{(m)} \tilde{p}^{(n)} d_{nm}^{(I)}\left(\tilde{\mathbf{R}}\right)$$

$$= \dot{P}_I + \hbar \sum_{n,m=1}^{F} \dot{\tilde{x}}^{(m)} \tilde{p}^{(n)} d_{nm}^{(I)}\left(\tilde{\mathbf{R}}\right) + \hbar \sum_{n,m=1}^{F} \tilde{x}^{(m)} \dot{\tilde{p}}^{(n)} d_{nm}^{(I)}\left(\tilde{\mathbf{R}}\right) + \hbar \sum_{n,m=1}^{F} \tilde{x}^{(m)} \tilde{p}^{(n)} \sum_{J=1}^{N} \frac{\partial d_{nm}^{(I)}\left(\tilde{\mathbf{R}}\right)}{\partial \tilde{R}_J} \dot{\tilde{R}}_J \quad . \quad (112)$$

Substitution of eqs (108)-(111) into eq (112) yields,

$$\dot{P}_I\left(\tilde{\mathbf{R}},\tilde{\mathbf{P}},\tilde{\mathbf{x}},\tilde{\mathbf{p}}\right) = \hbar \sum_{J=1}^{N} \sum_{n,m=1}^{F} \frac{P_J(\tilde{\mathbf{R}},\tilde{\mathbf{P}},\tilde{\mathbf{x}},\tilde{\mathbf{p}})}{M_J} \tilde{x}^{(n)} \tilde{p}^{(m)} \left( \frac{\partial d_{mn}^{(J)}\left(\tilde{\mathbf{R}}\right)}{\partial \tilde{R}_I} - \frac{\partial d_{mn}^{(I)}\left(\tilde{\mathbf{R}}\right)}{\partial \tilde{R}_J} \right)$$

$$- \sum_{k=1}^{F} \frac{\partial E_k\left(\tilde{\mathbf{R}}\right)}{\partial \tilde{R}_I} \left( \frac{1}{2} \left( \left(\tilde{x}^{(k)}\right)^2 + \left(\tilde{p}^{(k)}\right)^2 \right) - \gamma \right)$$

$$+ \hbar \sum_{n,m=1}^{F} \sum_{J=1}^{N} \sum_{k=1}^{F} \frac{P_J(\tilde{\mathbf{R}},\tilde{\mathbf{P}},\tilde{\mathbf{x}},\tilde{\mathbf{p}})}{M_J} d_{nk}^{(I)}\left(\tilde{\mathbf{R}}\right) d_{km}^{(J)}\left(\tilde{\mathbf{R}}\right) \left( \tilde{x}^{(m)} \tilde{p}^{(n)} - \tilde{x}^{(n)} \tilde{p}^{(m)} \right)$$

$$+ \sum_{n,m=1}^{F} \left( E_n\left(\tilde{\mathbf{R}}\right) \tilde{x}^{(n)} \tilde{x}^{(m)} - E_m\left(\tilde{\mathbf{R}}\right) \tilde{p}^{(m)} \tilde{p}^{(n)} \right) d_{nm}^{(I)}\left(\tilde{\mathbf{R}}\right) \quad . \quad (113)$$

Although $\dfrac{\partial d_{mn}^{(J)}\left(\tilde{\mathbf{R}}\right)}{\partial \tilde{R}_I} - \dfrac{\partial d_{mn}^{(I)}\left(\tilde{\mathbf{R}}\right)}{\partial \tilde{R}_J} = 0$ holds for $I \neq J$ for two-electronic-state systems as long as the time-reversal symmetry is satisfied, such an equality is, however, often not valid for $I \neq J$ for general nonadiabatic systems. Note that we have assumed that $\left\{ d_{mn}^{(I)}\left(\mathbf{R}\right) \right\}$ are real, which leads to $\mathbf{d}_{mn}^{(I)}\left(\mathbf{R}\right) = -\mathbf{d}_{nm}^{(I)}\left(\mathbf{R}\right)$. The third term of the right-hand side (RHS) of eq (113) becomes

$$\hbar \sum_{n,m=1}^{F} \sum_{J=1}^{N} \sum_{k=1}^{F} \frac{P_J(\tilde{\mathbf{R}},\tilde{\mathbf{P}},\tilde{\mathbf{x}},\tilde{\mathbf{p}})}{M_J} d_{nk}^{(I)}\left(\tilde{\mathbf{R}}\right) d_{km}^{(J)}\left(\tilde{\mathbf{R}}\right) \left( \tilde{x}^{(m)} \tilde{p}^{(n)} - \tilde{x}^{(n)} \tilde{p}^{(m)} \right)$$

$$= -\hbar \sum_{n,m=1}^{F} \sum_{J=1}^{N} \sum_{k=1}^{F} \frac{P_J(\tilde{\mathbf{R}},\tilde{\mathbf{P}},\tilde{\mathbf{x}},\tilde{\mathbf{p}})}{M_J} \left( d_{nk}^{(I)}\left(\tilde{\mathbf{R}}\right) d_{km}^{(J)}\left(\tilde{\mathbf{R}}\right) - d_{mk}^{(I)}\left(\tilde{\mathbf{R}}\right) d_{kn}^{(J)}\left(\tilde{\mathbf{R}}\right) \right) \tilde{x}^{(n)} \tilde{p}^{(m)} \quad . \quad (114)$$

$$= -\hbar \sum_{n,m=1}^{F} \sum_{J=1}^{N} \sum_{k=1}^{F} \frac{P_J(\tilde{\mathbf{R}},\tilde{\mathbf{P}},\tilde{\mathbf{x}},\tilde{\mathbf{p}})}{M_J} \left( d_{nk}^{(I)}\left(\tilde{\mathbf{R}}\right) d_{km}^{(J)}\left(\tilde{\mathbf{R}}\right) - d_{nk}^{(J)}\left(\tilde{\mathbf{R}}\right) d_{km}^{(I)}\left(\tilde{\mathbf{R}}\right) \right) \tilde{x}^{(n)} \tilde{p}^{(m)}$$

Substitution of eq (114) into eq (113) produces

$$\dot{P}_I\left(\tilde{\mathbf{R}}, \tilde{\mathbf{P}}, \tilde{\mathbf{x}}, \tilde{\mathbf{p}}\right) = \hbar \sum_{J=1}^{N} \sum_{n,m=1}^{F} \frac{P_J(\tilde{\mathbf{R}}, \tilde{\mathbf{P}}, \tilde{\mathbf{x}}, \tilde{\mathbf{p}})}{M_J} \tilde{x}^{(n)} \tilde{p}^{(m)} \left( \frac{\partial d_{mn}^{(J)}(\tilde{\mathbf{R}})}{\partial \tilde{R}_I} - \frac{\partial d_{mn}^{(I)}(\tilde{\mathbf{R}})}{\partial \tilde{R}_J} \right.$$
$$\left. - \sum_{k=1}^{F} \left( d_{nk}^{(I)}(\tilde{\mathbf{R}}) d_{km}^{(J)}(\tilde{\mathbf{R}}) - d_{nk}^{(J)}(\tilde{\mathbf{R}}) d_{km}^{(I)}(\tilde{\mathbf{R}}) \right) \right)$$
$$- \sum_{k=1}^{F} \frac{\partial E_k(\tilde{\mathbf{R}})}{\partial \tilde{R}_I} \left( \frac{1}{2}\left( \left(\tilde{x}^{(k)}\right)^2 + \left(\tilde{p}^{(k)}\right)^2 \right) - \gamma \right)$$
$$+ \sum_{n,m=1}^{F} \left( E_n(\tilde{\mathbf{R}}) \tilde{x}^{(n)} \tilde{x}^{(m)} - E_m(\tilde{\mathbf{R}}) \tilde{p}^{(m)} \tilde{p}^{(n)} \right) d_{nm}^{(I)}(\tilde{\mathbf{R}}) \quad . \tag{115}$$

The first term of the RHS of eq (115) indicates a relation to the non-abelian gauge theory, as we will demonstrate below.

The derivative of the first-order nonadiabatic coupling is

$$\frac{\partial d_{mn}^{(J)}}{\partial \tilde{R}_I} = \frac{\partial}{\partial \tilde{R}_I} \left\langle \phi_m \left| \frac{\partial}{\partial \tilde{R}_J} \phi_n \right\rangle \right. = \left\langle \frac{\partial}{\partial \tilde{R}_I} \phi_m \left| \frac{\partial}{\partial \tilde{R}_J} \phi_n \right\rangle \right. + \left\langle \phi_m \left| \frac{\partial^2}{\partial \tilde{R}_I \partial \tilde{R}_J} \phi_n \right\rangle \right. ,$$
$$= \sum_{k=1}^{F} d_{km}^{(I)} d_{kn}^{(J)} + D_{mn}^{(IJ)} \tag{116}$$

where $D_{mn}^{(IJ)} = \left\langle \phi_m \left| \frac{\partial^2}{\partial \tilde{R}_I \partial \tilde{R}_J} \phi_n \right\rangle \right.$ is a symmetric tensor of nuclear index, i.e., $D_{mn}^{(IJ)} = D_{mn}^{(JI)}$. It is then straightforward to show

$$\frac{\partial d_{mn}^{(J)}}{\partial \tilde{R}_I} - \frac{\partial d_{mn}^{(I)}}{\partial \tilde{R}_J} = -\sum_{k=1}^{F} \left( d_{mk}^{(I)} d_{kn}^{(J)} - d_{mk}^{(J)} d_{kn}^{(I)} \right) \quad , \tag{117}$$

which leads to the matrix form

$$\frac{\partial \mathbf{d}^{(J)}}{\partial \tilde{R}_I} - \frac{\partial \mathbf{d}^{(I)}}{\partial \tilde{R}_J} + [\mathbf{d}^{(I)}, \mathbf{d}^{(J)}]_{\text{ele}} = 0 \quad . \tag{118}$$

The LHS of eq (118) is an analogue to the gauge field tensor, a fundamental concept of non-abelian gauge field theory[307]. Consider vector $-i\mathbf{d}(\tilde{\mathbf{R}})$, of which the $J$-th component, $-i\mathbf{d}^{(J)}(\tilde{\mathbf{R}})$, is a Hermitian matrix of the electronic state DOFs (i.e., an $F \times F$ matrix). Vector $-i\mathbf{d}(\tilde{\mathbf{R}})$ is the non-abelian gauge potential[308], which is a generalization of the vector potential of the electromagnetic field. That is, each spacial component of the (magnetic) vector potential is generalized by a matrix of the inner DOFs (here the inner DOFs are those of the electronic states) for each nuclear DOF. Different from scalars, matrices often do not commute with each other, which is the origin of the 'non-abelian' characteristic. When there exists the diabatic representation of the system, the gauge field tensor,

$$\frac{\partial\left(-i\mathbf{d}^{(J)}\right)}{\partial \tilde{R}_I} - \frac{\partial\left(-i\mathbf{d}^{(I)}\right)}{\partial \tilde{R}_J} + i[-i\mathbf{d}^{(I)}, -i\mathbf{d}^{(J)}]_{\text{ele}} \quad , \quad (119)$$

is always zero when it is not in the region of conical intersections where some elements of $\{\partial \mathbf{d}^{(J)}/\partial \tilde{R}_I\}$ are not well-defined. (It is trivial to verify that eq (118) holds even when nonadiabatic coupling terms $\{\mathbf{d}_{mn}^{(I)}(\mathbf{R})\}$ are complex.)

We note that, eq (118) rigorously holds only when infinite electronic levels are considered for the general molecular system (i.e., the electronically diabatic representation is inherently well-defined). The LHS of eq (118) is often effectively zero as long as a finite discrete $F$-electronic-state description is reasonable for the molecular system.

Substitution of eq (117) into eq (115) yields

$$\begin{aligned}\dot{P}_I\left(\tilde{\mathbf{R}}, \tilde{\mathbf{P}}, \tilde{\mathbf{x}}, \tilde{\mathbf{p}}\right) = &-\sum_{k=1}^{F} \frac{\partial E_k\left(\tilde{\mathbf{R}}\right)}{\partial R_I}\left(\frac{1}{2}\left(\left(\tilde{x}^{(k)}\right)^2 + \left(\tilde{p}^{(k)}\right)^2\right) - \gamma\right) \\ &-\sum_{n,m=1}^{F}\left(E_n\left(\tilde{\mathbf{R}}\right) - E_m\left(\tilde{\mathbf{R}}\right)\right)d_{mn}^{(I)}\left(\tilde{\mathbf{R}}\right)\frac{1}{2}\left(\tilde{x}^{(n)}\tilde{x}^{(m)} + \tilde{p}^{(n)}\tilde{p}^{(m)}\right)\end{aligned} \quad , \quad (120)$$

which is equivalent to the second equation of eq 76 of the main text. It indicates that the canonical momentum in the diabatic representation is covariant with *kinematic* momentum $\mathbf{P}$ rather than canonical momentum $\tilde{\mathbf{P}}$ in adiabatic representation unless all nonadiabatic coupling terms vanish. This is consistent with the spirit of the work of Cotton *et. al.* in ref [186]. Because the EOMs (eq 70 and eq 76) of the main text are identical to Hamilton's EOMs generated by eq (107), the mapping Hamiltonian (eq 79 or eq (107)) is conserved during the evolution in the adiabatic representation.

It is straightforward to extend the discussion to the case when nonadiabatic coupling terms $\{\mathbf{d}_{mn}^{(I)}(\mathbf{R})\}$ are complex. The conclusion is similar.

**Appendix 3. The relationship between constraint coordinate-momentum phase space and Stratonovich phase space**

Stratonovich's original work[100] in 1956 maps a 2-state (spin-1/2) system onto a two-dimensional sphere. We review two kinds of further developments of the Stratonovich-Weyl mapping phase space representations for $F$-state quantum system: the first one based on the $\mathrm{SU}(2)$ structure[103, 113] and the second one based on the $\mathrm{SU}(F)$ structure[114]. We show the relationship between constrained coordinate-momentum phase space representation that we use in the Focus Article and the two kinds of Stratonovich phase space representations.

In the $\mathrm{SU}(2)$ Stratonovich phase space representation, an $F$-state system is treated as a spin-$j$ system (where $F = 2j+1$). The basis set consists of $|j,m\rangle$, the eigenstate of the square of total angular momentum $\hat{J}^2$ and the $z$-component of angular momentum $\hat{J}_z$ with quantum numbers $j$ and $m$, respectively. The mapping kernel is

$$\hat{K}_{\mathrm{ele}}^{\mathrm{SU}(2)}(\theta,\varphi;s) = \sqrt{\frac{\pi}{2j+1}} \sum_{l=0}^{2j} (C_{jj,l0}^{jj})^{-s} \sum_{m=-l}^{l} Y_{lm}^*(\theta,\varphi)\hat{T}_{lm}^j, \quad s \in \mathbb{R} \tag{121}$$

where $\hat{T}_{lm}^j$ is the irreducible tensor operator defined as[131]

$$\hat{T}_{lm}^j = \sqrt{\frac{2l+1}{2j+1}} \sum_{m',n=-j}^{j} C_{jm',lm}^{jn} |j,n\rangle\langle j,m'|. \tag{122}$$

Here $C_{j_1m_1,j_2m_2}^{jm} = \langle jm | j_1m_1, j_2m_2 \rangle$ is the well-known Clebsch-Gordan coefficient for the angular momentum coupling, and $Y_{lm}(\theta,\varphi)$ is the spherical harmonic function. The inverse kernel of $\hat{K}_{\mathrm{ele}}^{\mathrm{SU}(2)}(\theta,\varphi;s)$ is simply $\hat{K}_{\mathrm{ele}}^{\mathrm{SU}(2)}(\theta,\varphi;-s)$. We note that, although $s=1, 0$, and $-1$ are traditionally associated with the $Q$, Wigner, and $P$-functions respectively and used in the literature[111, 128], parameter $s$ of eq (121) can in principle take *any* real value.

When $F > 2$, the $\mathrm{SU}(2)$ Stratonovich phase space $(\theta, \varphi)$ does *not* have a phase point-to-phase point mapping to constraint coordinate-momentum phase space, although the relation can only be constructed by virtue of the density matrix. Only when $F = 2$ as in the original work of Stratonovich, there exists a phase point-to-phase point mapping to constraint coordinate-momentum phase space. The mapping kernel can be expressed in terms of the spin-coherent state,

$$\hat{K}_{\mathrm{ele}}^{\mathrm{SU}(2)}(\theta, \varphi; s) = 3^{(1+s)/2} |\theta, \varphi\rangle\langle\theta, \varphi| + \frac{\hat{\mathbf{I}}}{2}\left[1 - 3^{(1+s)/2}\right] . \quad (123)$$

In eq (123) spin coherent state $|\theta, \varphi\rangle$ is

$$|\theta, \varphi\rangle = \begin{pmatrix} e^{-i\varphi} \sin(\theta/2) \\ \cos(\theta/2) \end{pmatrix}, \quad (124)$$

where $(\theta, \varphi)$ are the spherical coordinate variables on the two-dimensional spherical phase space. The range of $\theta$ is $[0, \pi]$ and that for $\varphi$ is $[0, 2\pi)$. When $F = 2$, the explicit transformation of $(\theta, \varphi)$ to the constrained coordinate-momentum phase space $(x^{(1)}, x^{(2)}, p^{(1)}, p^{(2)})$ is

$$\begin{pmatrix} x^{(1)} \\ p^{(1)} \end{pmatrix} = \sqrt{2(1+2\gamma)} \begin{pmatrix} \cos\psi & -\sin\psi \\ \sin\psi & \cos\psi \end{pmatrix} \begin{pmatrix} \cos\varphi \sin(\theta/2) \\ -\sin\varphi \sin(\theta/2) \end{pmatrix}$$
$$\begin{pmatrix} x^{(2)} \\ p^{(2)} \end{pmatrix} = \sqrt{2(1+2\gamma)} \begin{pmatrix} \cos\psi & -\sin\psi \\ \sin\psi & \cos\psi \end{pmatrix} \begin{pmatrix} \cos(\theta/2) \\ 0 \end{pmatrix} \quad (125)$$

where $\psi$ is an additional global phase.

The second kind of representation is the $\mathrm{SU}(F)$ Stratonovich phase space. A kind as described by Tilma *et.al.* in ref [114] is diffeomorphic to the quotient set $\mathrm{SU}(F)/\mathrm{U}(F-1)$, parameterized by $(2F-2)$ angle variables $(\boldsymbol{\theta}, \boldsymbol{\varphi}) = (\theta_1, \theta_2, \ldots, \theta_{F-1}, \varphi_1, \varphi_2, \ldots, \varphi_{F-1})$. The range of each angle $\theta_i$ is $[0, \pi/2]$ and that for each angle $\varphi_i$ is $[0, 2\pi)$. The $\mathrm{SU}(F)$ Stratonovich phase space of ref [114] has been used to prepare the initial condition for the Meyer-Miller mapping model of non-adiabatic dynamics in ref [199].

The mapping kernel of the $\mathrm{SU}(F)/\mathrm{U}(F-1)$ Stratonovich phase space of ref [114] is

$$\hat{K}_{\text{ele}}^{\text{SU}(F)}(\boldsymbol{\theta},\boldsymbol{\varphi};s) = (1+F)^{(1+s)/2}\,|\boldsymbol{\theta},\boldsymbol{\varphi}\rangle\langle\boldsymbol{\theta},\boldsymbol{\varphi}| + \frac{\hat{\mathbf{I}}}{F}\left(1-(1+F)^{(1+s)/2}\right), \quad s \in \mathbb{R} \tag{126}$$

and the inverse kernel is simply $\hat{K}_{\text{ele}}^{\text{SU}(F)}(\boldsymbol{\theta},\boldsymbol{\varphi};-s)$. The explicit form of the generalized coherent state $|\boldsymbol{\theta},\boldsymbol{\varphi}\rangle$ is[114,242,243]

$$|\boldsymbol{\theta},\boldsymbol{\varphi}\rangle = \sum_{n=1}^{F} c_n\,|n\rangle\ , \tag{127}$$

where the coefficients are

$$\begin{pmatrix} c_1 \\ c_2 \\ c_3 \\ \vdots \\ c_{F-3} \\ c_{F-2} \\ c_{F-1} \\ c_F \end{pmatrix} = \begin{pmatrix} e^{i(\varphi_1+\varphi_2+\cdots+\varphi_{F-1})}\cos(\theta_1)\cos(\theta_2)\cdots\cos(\theta_{F-2})\sin(\theta_{F-1}) \\ -e^{i(-\varphi_1+\varphi_2+\cdots+\varphi_{F-1})}\sin(\theta_1)\cos(\theta_2)\cdots\cos(\theta_{F-2})\sin(\theta_{F-1}) \\ -e^{i(\varphi_3+\varphi_4+\cdots+\varphi_{F-1})}\sin(\theta_2)\cos(\theta_3)\cdots\cos(\theta_{F-2})\sin(\theta_{F-1}) \\ \vdots \\ -e^{i(\varphi_{F-3}+\varphi_{F-2}+\varphi_{F-1})}\sin(\theta_{F-4})\cos(\theta_{F-3})\cos(\theta_{F-2})\sin(\theta_{F-1}) \\ -e^{i(\varphi_{F-2}+\varphi_{F-1})}\sin(\theta_{F-3})\cos(\theta_{F-2})\sin(\theta_{F-1}) \\ -e^{i(\varphi_{F-1})}\sin(\theta_{F-2})\sin(\theta_{F-1}) \\ \cos(\theta_{F-1}) \end{pmatrix}. \tag{128}$$

As derived first in Appendix A of ref [134] in the spirit of ref [133] and then in the Supporting Information of ref [136], the mapping kernel of constraint coordinate-momentum phase space for a set of $F$ states (eq 29 of the main text) is denoted as,

$$\hat{K}_{\text{ele}}(\mathbf{x},\mathbf{p};\gamma) = \sum_{m,n=1}^{F}\left[\frac{(x^{(m)}-\mathrm{i}p^{(m)})(x^{(n)}+\mathrm{i}p^{(n)})}{2} - \gamma\delta_{mn}\right]|n\rangle\langle m| = |\mathbf{x},\mathbf{p}\rangle\langle\mathbf{x},\mathbf{p}| - \gamma\hat{\mathbf{I}}, \tag{129}$$

where the non-normalized state $|\mathbf{x},\mathbf{p}\rangle$ is

$$\sum_{n=1}^{F}\frac{x^{(n)}+\mathrm{i}p^{(n)}}{\sqrt{2}}\,|n\rangle. \tag{130}$$

The expression of eq (130) for the amplitudes of being in different states was already used in Appendix B of ref [133], while the action-angle version of eq (130) was earlier presented in Meyer and Miller's seminal paper[155]. The $\mathrm{U}(F)$ constraint coordinate-momentum phase space is diffeomorphic to $\mathrm{U}(F)/\mathrm{U}(F-1)$, which is equivalent to $SU(F)/SU(F-1)$ [309] because both of them lead to the $(2F-1)$-dimensional sphere in $2F$ dimensional Euclidean space[310, 311]. Comparison of $\hat{K}_{\text{ele}}^{SU(F)}(\boldsymbol{\theta},\boldsymbol{\varphi};s)$ to $\hat{K}_{\text{ele}}(\mathbf{x},\mathbf{p};\gamma)$ implies that the two kernels are closely related. The correspondence between parameters $s$ and $\gamma$ reads

$$1 + F\gamma = (1+F)^{(1+s)/2} \quad . \tag{131}$$

It is evident (from eq (131), the relation between constraint coordinate-momentum phase space and the $SU(F)/U(F-1)$ Stratonovich phase space) that parameter $s$ can be any real number in eq (126), *not* limited to $s = 1, 0$, or $-1$ of refs [114, 115, 117, 199], where the corresponding value of parameter $\gamma$ of constraint coordinate-momentum phase space[133, 134, 136] is $\gamma = 1, (\sqrt{1+F}-1)/F$, or $0$. This has been clearly mentioned in refs [57, 58, 136].

Any normalized pure state of the $F$-dimensional Hilbert space uniquely corresponds to a state on constraint coordinate-momentum phase space, $|\mathbf{x},\mathbf{p}\rangle$, which is equivalent to $\exp[i\psi]|\boldsymbol{\theta},\boldsymbol{\varphi}\rangle$ with $\psi$ as the global phase. That is, the correspondence between $(\boldsymbol{\theta},\boldsymbol{\varphi})$ and $(\mathbf{x},\mathbf{p})$ is

$$\begin{pmatrix} x^{(n)} \\ p^{(n)} \end{pmatrix} = \sqrt{2(1+F\gamma)} \begin{pmatrix} \cos\psi & -\sin\psi \\ \sin\psi & \cos\psi \end{pmatrix} \begin{pmatrix} \text{Re}\langle n|\boldsymbol{\theta},\boldsymbol{\varphi}\rangle \\ \text{Im}\langle n|\boldsymbol{\theta},\boldsymbol{\varphi}\rangle \end{pmatrix} \quad . \tag{132}$$

Under the transformation, eq (132), mapping functions $A_C(\boldsymbol{\theta},\boldsymbol{\varphi};s) = \text{Tr}\left[\hat{A}\,\hat{K}_{\text{ele}}^{SU(F)}(\boldsymbol{\theta},\boldsymbol{\varphi};s)\right]$ and $A_C(\mathbf{x},\mathbf{p};\gamma) = \text{Tr}\left[\hat{A}\,\hat{K}_{\text{ele}}(\mathbf{x},\mathbf{p};\gamma)\right]$ of an operator $\hat{A}$ share the same value. The global phase, $\psi$, which is missing in the $SU(F)/U(F-1)$ Stratonovich phase space[114], however, is important for the expression of quantum dynamics as linear equations of motion.

When we consider quantum dynamics in a finite $F$-dimensional Hilbert space, if the Hamiltonian operator includes linear components beyond the identity operator and generator operators of phase space group, it is impossible to derive trajectory-based exact dynamics. The $SU(2)$ group involves the identity operator and angular momentum operators as generators on $\mathbb{S}^2$ sphere. It produces trajectory-based exact dynamics only for two-state systems, but fails to do so for all $F > 2$ cases. It is claimed that trajectory-based dynamics is a good approximation for the large spin limit ($F \to \infty$) though [109, 111, 132]. Except for the $F=2$ case, the expression of quantum dynamics on the $SU(2)$ Stratonovich phase space has no direct relation to the trajectory-based exact dynamics on constrained coordinate-momentum phase space.

The $SU(F)$ Stratonovich phase space, however, produces trajectory-based exact dynamics for the finite $F$-dimensional Hilbert space. This is because that the evolution generated by any Hamiltonian of the $F$-dimensional Hilbert space is the action of some group elements of $SU(F)$.[304] The inherent symplectic structure of $SU(F)/U(F-1)$ Stratonovich phase space [312] indicates that the trajectory-based exact dynamics can be produced by the corresponding mapping Hamiltonian function $H_C = \text{Tr}[\hat{H}\hat{K}_{\text{ele}}^{\text{SU}(F)}(\boldsymbol{\theta}, \boldsymbol{\varphi}; s)]$, i.e.,

$$\begin{cases} \dot{\theta}_i = -\sum_{j=1}^{F-1} \frac{A_{ji}}{(1+F)^{(1+s)/2}} \frac{\partial H_C}{\partial \varphi_j} \\ \dot{\varphi}_i = +\sum_{j=1}^{F-1} \frac{A_{ij}}{(1+F)^{(1+s)/2}} \frac{\partial H_C}{\partial \theta_j} \end{cases} \tag{133}$$

The elements, $\{A_{ij}\}$, of the $(F-1)\times(F-1)$ matrix, $\mathbf{A}$, are

$$A_{ij} = \begin{cases} \frac{1}{2}\csc(2\theta_1)\csc^2\theta_{F-1}\prod_{k=2}^{F-2}\sec^2\theta_k, & i = j = 1; \\ -\frac{1}{2}\cot(2\theta_1)\csc^2\theta_{F-1}\prod_{k=2}^{F-2}\sec^2\theta_k, & (i-1) = j = 1; \\ \csc(2\theta_i)\csc^2\theta_{F-1}\prod_{k=i+1}^{F-2}\sec^2\theta_k, & 2 \leq i = j \leq F-2; \\ -\frac{1}{2}\cot(\theta_i)\csc^2\theta_{F-1}\prod_{k=i+1}^{F-2}\sec^2\theta_k, & 2 \leq (i-1) = j \leq F-2; \\ -\csc(2\theta_{F-1}), & i = j = (F-1). \end{cases} \tag{134}$$

The explicit expression of the EOMs of eq (133) is, however, much complicated. In addition, because trigonometric functions are involved in eq (134), singularities are inevitable in the EOMs on the $SU(F)/U(F-1)$ Stratonovich phase space for $F > 2$. This makes the expression of quantum dynamics on the $SU(F)/U(F-1)$ Stratonovich phase space numerically unfavorable. In comparison, on (weighted) constrained coordinate-momentum phase space, Hamilton's EOMs are simply linear in derivatives with coefficients independent of phase variables, as well as exact.

The relation of eq (132) implies the subtle difference between exact trajectory-base dynamics on the $\mathrm{SU}(F)/\mathrm{U}(F-1)$ Stratonovich phase space and that on constraint coordinate-momentum phase space. The addition of the global phase, $\psi$, is critical to obtain a one-to-one correspondence between the two approaches. The EOM of $\psi$ reads

$$\dot{\psi} = \left[ -\frac{\partial}{\partial \lambda} + \frac{\tan \theta_{F-1}}{2\lambda} \frac{\partial}{\partial \theta_{F-1}} \right] H_C^{(\lambda)}(\boldsymbol{\theta}, \boldsymbol{\varphi}; s) \bigg|_{\lambda = (1+F)^{\frac{1+s}{2}}}, \quad (135)$$

where the extended mapping Hamiltonian function is defined by

$$H_C^{(\lambda)}(\boldsymbol{\theta}, \boldsymbol{\varphi}; s) = \mathrm{Tr}\left[ \hat{K}_{\mathrm{ele}}^{\mathrm{SU}(F),(\lambda)}(\boldsymbol{\theta}, \boldsymbol{\varphi}; s) \hat{H} \right] \quad (136)$$

for the extended $\mathrm{SU}(F)/\mathrm{U}(F-1)$ Stratonovich mapping 'kernel'

$$\hat{K}_{\mathrm{ele}}^{\mathrm{SU}(F),(\lambda)}(\boldsymbol{\theta}, \boldsymbol{\varphi}; s) = \lambda |\boldsymbol{\theta}, \boldsymbol{\varphi}\rangle\langle\boldsymbol{\theta}, \boldsymbol{\varphi}| + \frac{\hat{I}}{F}\left( 1 - (1+F)^{\frac{1+s}{2}} \right). \quad (137)$$

In eq (137) $\lambda$ is treated as an 'invariant' variable. The evolution of state $\sqrt{\lambda} e^{i\psi} |\boldsymbol{\theta}, \boldsymbol{\varphi}\rangle \equiv |\mathbf{x}, \mathbf{p}\rangle$ generates

$$\begin{cases} \dot{\lambda} = \dfrac{\partial}{\partial \psi} H_C^{(\lambda)}(\boldsymbol{\theta}, \boldsymbol{\varphi}; s) = 0 \\[6pt] \dot{\psi} = \left[ -\dfrac{\partial}{\partial \lambda} + \dfrac{\tan \theta_{F-1}}{2\lambda} \dfrac{\partial}{\partial \theta_{F-1}} \right] H_C^{(\lambda)}(\boldsymbol{\theta}, \boldsymbol{\varphi}; s) \\[6pt] \dot{\theta}_i = -\displaystyle\sum_{j=1}^{F-1} \dfrac{A_{ji}}{\lambda} \dfrac{\partial}{\partial \varphi_j} H_C^{(\lambda)}(\boldsymbol{\theta}, \boldsymbol{\varphi}; s) - \delta_{i,F-1} \dfrac{\tan \theta_{F-1}}{2\lambda} \dfrac{\partial}{\partial \psi} H_C^{(\lambda)}(\boldsymbol{\theta}, \boldsymbol{\varphi}; s) \\[6pt] \dot{\varphi}_i = +\displaystyle\sum_{j=1}^{F-1} \dfrac{A_{ij}}{\lambda} \dfrac{\partial}{\partial \theta_j} H_C^{(\lambda)}(\boldsymbol{\theta}, \boldsymbol{\varphi}; s) \end{cases} \quad (138)$$

It is straightforward to show that under the bijection,

$$\begin{pmatrix} x^{(n)} \\ p^{(n)} \end{pmatrix} = \sqrt{2\lambda} \begin{pmatrix} \cos\psi & -\sin\psi \\ \sin\psi & \cos\psi \end{pmatrix} \begin{pmatrix} \mathrm{Re}\langle n|\boldsymbol{\theta}, \boldsymbol{\varphi}\rangle \\ \mathrm{Im}\langle n|\boldsymbol{\theta}, \boldsymbol{\varphi}\rangle \end{pmatrix}, \quad (139)$$

between variables $(\lambda, \psi, \boldsymbol{\theta}, \boldsymbol{\varphi})$ and $(\mathbf{x}, \mathbf{p})$, eq (138) leads to the EOMs of phase variables of (weighted) constraint phase space

$$\dot{x}^{(n)} = +\frac{\partial H_C}{\partial p^{(n)}}$$
$$\dot{p}^{(n)} = -\frac{\partial H_C}{\partial x^{(n)}} \quad (140)$$

by the chain rules, i.e.,

$$\dot{x}^{(n)} = \sum_{i=1}^{2F} \frac{\partial x^{(n)}}{\partial z_i} \dot{z}_i$$
$$\dot{p}^{(n)} = \sum_{i=1}^{2F} \frac{\partial p^{(n)}}{\partial z_i} \dot{z}_i \quad (141)$$

and

$$\frac{\partial H_C}{\partial z_i} = \sum_{n=1}^{F} \frac{\partial x^{(n)}}{\partial z_i} \frac{\partial H_C}{\partial x^{(n)}} + \sum_{n=1}^{F} \frac{\partial p^{(n)}}{\partial z_i} \frac{\partial H_C}{\partial p^{(n)}} \quad , \quad (142)$$

where $\{z_i\}$ denote variables $\{\lambda, \psi, \boldsymbol{\theta}, \boldsymbol{\varphi}\}$.

The relation has already been clearly pointed out in refs [57, 58]. In this Appendix, we demonstrate the subtle difference between the nonlinear EOMs of the mapping Hamiltonian on the $\mathrm{SU}(F)/\mathrm{U}(F-1)$ Stratonovich phase space and the linear EOMs of the mapping Hamiltonian on the $\mathrm{U}(F)/\mathrm{U}(F-1)$ constraint coordinate-momentum phase space. Without the definition of the 'invariant' variable, $\lambda$, in the 'extended' $\mathrm{SU}(F)/\mathrm{U}(F-1)$ Stratonovich mapping kernel and the EOM of the global phase, $\psi$, (i.e., eq (135) and eq (137)), it is *impossible* to rigorously derive the trajectory-based dynamics generated by the Meyer-Miller mapping Hamiltonian. In comparison, the EOMs of Meyer-Miller mapping Hamiltonian (with $2F$ variables) is naturally derived in constraint coordinate-momentum phase space of refs [57, 58, 133, 134, 136] for the discrete $F$-state system.

Lang *et al.* have recently directly employed $F^2-1$ variables of $\mathrm{SU}(F)$ in the EOMs of the discrete $F$-state sytems[60]. It is straightforward to show that such an approach can be included in our comprehensive phase space mapping theory with the commutator matrix[57, 58], where it is easy to obtain the manifold structure.

**Appendix 4. Marginal distribution functions on symmetrically weighted coordinate-momentum phase space**

As demonstrated in Figure 5, the marginal distribution functions of a spin-1/2 system on symmetrically weighted constraint coordinate-momentum phase space demonstrate a hollow structure. Equation 50 leads to the marginal functions on symmetrically weighted coordinate-momentum phase space

$$\begin{aligned}
\mathcal{K}_{\uparrow\uparrow}(x^{(1)}, x^{(2)}) &= \frac{1-2\Delta^2+2\Delta}{4\Delta} \frac{1+(x^{(1)})^2/2-(x^{(2)})^2/2}{2\pi(1+2\Delta)}\bigg|_{(x^{(1)})^2+(x^{(2)})^2 \leq 2(1+2\Delta)} \\
&\quad - \frac{1-2\Delta^2-2\Delta}{4\Delta} \frac{1+(x^{(1)})^2/2-(x^{(2)})^2/2}{2\pi(1-2\Delta)}\bigg|_{(x^{(1)})^2+(x^{(2)})^2 \leq 2(1-2\Delta)} \\
\mathcal{K}_{\uparrow\downarrow}(x^{(1)}, x^{(2)}) &= \mathcal{K}_{\downarrow\uparrow}(x^{(1)}, x^{(2)}) = \frac{1-2\Delta^2+2\Delta}{4\Delta} \frac{x^{(1)}x^{(2)}}{2\pi(1+2\Delta)}\bigg|_{(x^{(1)})^2+(x^{(2)})^2 \leq 2(1+2\Delta)} \\
&\quad - \frac{1-2\Delta^2-2\Delta}{4\Delta} \frac{x^{(1)}x^{(2)}}{2\pi(1-2\Delta)}\bigg|_{(x^{(1)})^2+(x^{(2)})^2 \leq 2(1-2\Delta)} \\
\mathcal{K}_{\downarrow\downarrow}(x^{(1)}, x^{(2)}) &= \frac{1-2\Delta^2+2\Delta}{4\Delta} \frac{1-(x^{(1)})^2/2+(x^{(2)})^2/2}{2\pi(1+2\Delta)}\bigg|_{(x^{(1)})^2+(x^{(2)})^2 \leq 2(1+2\Delta)} \\
&\quad - \frac{1-2\Delta^2-2\Delta}{4\Delta} \frac{1-(x^{(1)})^2/2+(x^{(2)})^2/2}{2\pi(1-2\Delta)}\bigg|_{(x^{(1)})^2+(x^{(2)})^2 \leq 2(1-2\Delta)}
\end{aligned} \quad (143)$$

As $\Delta \to 0^+$, $\mathcal{K}_{nm}(x^{(1)}, x^{(2)})$ approaches zero in region $(x^{(1)})^2+(x^{(2)})^2 \leq 2(1-2\Delta)$, yielding the hollow structure. We note that the hollow structure appears only for $F=2$ but not for $F \geq 3$.

**Appendix 5. The Wigner-Moyal equation of composite systems and its semiclassical limit**

The quantum Liouville theorem or the von Neumann equation reads

$$\frac{\partial}{\partial t}\hat{\rho} = -\frac{1}{i\hbar}\left[\hat{\rho}, \hat{H}\right] \quad . \tag{144}$$

Its expression on quantum phase space leads to the (general) Wigner-Moyal equation,

$$\frac{\partial}{\partial t}\rho(\mathbf{X}) = -\{\{\rho(\mathbf{X}), H(\mathbf{X})\}\} = -\frac{1}{i\hbar}\left(\rho(\mathbf{X}) \star H(\mathbf{X}) - H(\mathbf{X}) \star \rho(\mathbf{X})\right). \tag{145}$$

Here $\star$ represents the general Moyal product[94] operation on phase space functions, and $\{\{A, B\}\} = \frac{1}{i\hbar}(A \star B - B \star A)$ denotes the general Moyal bracket.

In the continuous-variable quantum system, only nuclear mapping DOFs $\mathbf{X} = (\mathbf{R}, \mathbf{P})$ are involved, and the Moyal product[37, 39, 44, 90, 91, 94] reads

$$\star\big|_{\mathrm{nuc}} = \exp\left[i\frac{\hbar}{2}\left(\overleftarrow{\nabla}_{\mathbf{R}} \cdot \overrightarrow{\nabla}_{\mathbf{P}} - \overleftarrow{\nabla}_{\mathbf{P}} \cdot \overrightarrow{\nabla}_{\mathbf{R}}\right)\right] . \tag{146}$$

When the discrete-variable quantum system is studied, only electronic mapping DOFs $\mathbf{X} = (\mathbf{x}, \mathbf{p})$ are included, because all phase space functions are quadratic in the constraint coordinate-momentum phase space representation, the general Moyal product involves a cut-off formalism,

$$\star\big|_{\mathrm{ele}} = \hat{\Xi} - \gamma\hat{\Xi}\hat{\Xi}^*, \tag{147}$$

where

$$\begin{aligned}\hat{\Xi} &= \frac{1}{2}\left(\overleftarrow{\nabla}_{\mathbf{x}} - i\overleftarrow{\nabla}_{\mathbf{p}}\right) \cdot \left(\overrightarrow{\nabla}_{\mathbf{x}} + i\overrightarrow{\nabla}_{\mathbf{p}}\right) \\ &= \sum_{n=1}^{F}\frac{1}{2}\left(\overleftarrow{\frac{\partial}{\partial x_n}} - i\overleftarrow{\frac{\partial}{\partial p_n}}\right)\left(\overrightarrow{\frac{\partial}{\partial x_n}} + i\overrightarrow{\frac{\partial}{\partial p_n}}\right)\end{aligned} . \tag{148}$$

The arrows above the gradient operators in eq (146) or eq (148) indicate the directions in which they act. When we consider the composite system, the total general Moyal product is expressed as the direct product

$$\star = \star\big|_{\mathrm{nuc}} \otimes \star\big|_{\mathrm{ele}} \tag{149}$$

for hybrid phase variables $\mathbf{X} = (\mathbf{R}, \mathbf{P}, \mathbf{x}, \mathbf{p})$ in the diabatic representation. When only the terms up to the first-order of $\hbar$ of eq (146) are kept (similar to the mixed quantum-classical approach[53]), the corresponding Moyal product of eq (149) becomes

$$\star = \left(1 + i\frac{\hbar}{2}\left(\overleftarrow{\nabla}_{\mathbf{R}} \cdot \overrightarrow{\nabla}_{\mathbf{P}} - \overleftarrow{\nabla}_{\mathbf{P}} \cdot \overrightarrow{\nabla}_{\mathbf{R}}\right)\right) \otimes \left(\hat{\Xi} - \gamma\hat{\Xi}\hat{\Xi}^*\right) + O\left(\hbar^2\right). \tag{150}$$

and the generalized Wigner-Moyal equation of eq (145) then reads

$$\begin{aligned}
\frac{\partial}{\partial t}\rho(\mathbf{X}) &= -\frac{1}{i\hbar}\left(\rho(\mathbf{X})\hat{\Xi}H(\mathbf{X}) - H(\mathbf{X})\hat{\Xi}\rho(\mathbf{X})\right) \\
&\quad - \frac{1}{i\hbar}\left(\rho(\mathbf{X})\left(i\frac{\hbar}{2}\left(\overleftarrow{\nabla}_{\mathbf{R}} \cdot \overrightarrow{\nabla}_{\mathbf{P}} - \overleftarrow{\nabla}_{\mathbf{P}} \cdot \overrightarrow{\nabla}_{\mathbf{R}}\right)\right) \otimes \left(\hat{\Xi} - \gamma\hat{\Xi}\hat{\Xi}^*\right)H(\mathbf{X})\right. \\
&\quad \left. - H(\mathbf{X})\left(i\frac{\hbar}{2}\left(\overleftarrow{\nabla}_{\mathbf{R}} \cdot \overrightarrow{\nabla}_{\mathbf{P}} - \overleftarrow{\nabla}_{\mathbf{P}} \cdot \overrightarrow{\nabla}_{\mathbf{R}}\right)\right) \otimes \left(\hat{\Xi} - \gamma\hat{\Xi}\hat{\Xi}^*\right)\rho(\mathbf{X})\right) \\
&= -\frac{1}{i\hbar}\left(\rho(\mathbf{X})\hat{\Xi}H(\mathbf{X}) - H(\mathbf{X})\hat{\Xi}\rho(\mathbf{X})\right) \\
&\quad - \frac{1}{2}\left(\nabla_{\mathbf{R}}\rho(\mathbf{X})\left(\hat{\Xi} - \gamma\hat{\Xi}\hat{\Xi}^*\right)\nabla_{\mathbf{P}}H(\mathbf{X}) - \nabla_{\mathbf{P}}\rho(\mathbf{X})\left(\hat{\Xi} - \gamma\hat{\Xi}\hat{\Xi}^*\right)\nabla_{\mathbf{R}}H(\mathbf{X})\right. \\
&\quad \left. - \nabla_{\mathbf{R}}H(\mathbf{X})\left(\hat{\Xi} - \gamma\hat{\Xi}\hat{\Xi}^*\right)\nabla_{\mathbf{P}}\rho(\mathbf{X}) + \nabla_{\mathbf{P}}H(\mathbf{X})\left(\hat{\Xi} - \gamma\hat{\Xi}\hat{\Xi}^*\right)\nabla_{\mathbf{R}}\rho(\mathbf{X})\right)
\end{aligned} \tag{151}$$

A simplified version of eq (151) reads

$$\begin{aligned}
-\frac{\partial}{\partial t}\rho(\mathbf{X}) &= \frac{1}{\hbar}\left(\nabla_{\mathbf{x}}\rho(\mathbf{X}) \cdot \nabla_{\mathbf{p}}H(\mathbf{X}) - \nabla_{\mathbf{x}}H(\mathbf{X}) \cdot \nabla_{\mathbf{p}}\rho(\mathbf{X})\right) \\
&\quad + \nabla_{\mathbf{R}}\rho(\mathbf{X}) \cdot \nabla_{\mathbf{P}}H(\mathbf{X}) \\
&\quad - \frac{1}{2}\left(\nabla_{\mathbf{P}}\rho(\mathbf{X})\left(\hat{\Xi} - \gamma\hat{\Xi}\hat{\Xi}^*\right)\nabla_{\mathbf{R}}H(\mathbf{X}) + \nabla_{\mathbf{R}}H(\mathbf{X})\left(\hat{\Xi} - \gamma\hat{\Xi}\hat{\Xi}^*\right)\nabla_{\mathbf{P}}\rho(\mathbf{X})\right)
\end{aligned} \tag{152}$$

Equation (152) is the expression of the mixed quantum-classical limit where the limit $\hbar \to 0$ is taken for only nuclear DOFs.

When dynamics of $\mathbf{X} = (\mathbf{R}, \mathbf{P}, \mathbf{x}, \mathbf{p})$ are approximated by the EOMs generated by the mapping Hamiltonian[57, 58], or the linearized semiclassical initial value representation (LSC-IVR) for only nuclear DOFs as first introduced in the constraint coordinate-momentum phase space representation in refs [134, 136], the exact generalized Wigner-Moyal equation of eq (149) is replaced by the Poisson bracket

$$\begin{aligned}
\{\rho(\mathbf{X}), H(\mathbf{X})\}_{Poisson} &= \frac{1}{\hbar}\left(\nabla_{\mathbf{x}}\rho(\mathbf{X}) \cdot \nabla_{\mathbf{p}}H(\mathbf{X}) - \nabla_{\mathbf{x}}H(\mathbf{X}) \cdot \nabla_{\mathbf{p}}\rho(\mathbf{X})\right) \\
&\quad + \left(\nabla_{\mathbf{R}}\rho(\mathbf{X}) \cdot \nabla_{\mathbf{P}}H(\mathbf{X}) - \nabla_{\mathbf{P}}\rho(\mathbf{X}) \cdot \nabla_{\mathbf{R}}H(\mathbf{X})\right)
\end{aligned} \tag{153}$$

While the first term of the RHS of eq (152) is equivalent to the first term of eq (153), the Poisson bracket of electronic mapping variables, the sum of the second and third terms of eq (152) are *not* exactly the same as the second term of eq (153), the Poisson bracket of nuclear DOFs.  It indicate that the EOMs governed by the mapping Hamiltonian[57, 58, 134, 136], are *not* the mixed quantum-classical limit of exact nonadiabatic dynamics, despite that the mapping Hamiltonian generates exact linear EOMs for composite (nonadiabatic) quantum systems in the frozen-nuclei limit (i.e., all terms that involve the derivatives of nuclear DOFs vanish).